\documentclass[11pt,a4paper]{article}
\usepackage{jheppub}
\usepackage{latexsym}
\usepackage{pdflscape}
\usepackage{bm}
\numberwithin{equation}{section}

\allowdisplaybreaks[2]

\def\cO{\mathcal{O}}
\def\cJ{\mathcal{J}}
\def\cD{\mathcal{D}}
\def\cN{\mathcal{N}}

\def\cF{\mathcal{F}}

\def\mint{\int_{-\infty}^\infty\!\cdots\!\int_{-\infty}^\infty}

\newcommand{\be}{\begin{equation}}
\newcommand{\ee}{\end{equation}}
\newcommand{\ba}{\begin{aligned}}
\newcommand{\ea}{\end{aligned}}

\def\Res{\mathop {\rm Res} \limits}

\DeclareMathOperator{\arcsinh}{arcsinh}

\DeclareMathOperator{\arccoth}{arccoth}

\def\bra#1{\left\langle #1 \right|}
\def\ket#1{\left| #1 \right\rangle}

\DeclareMathOperator{\Li}{Li}

\DeclareMathOperator{\B}{B}

\def\({\left(}
\def\){\right)}

\newcommand{\pd}{\partial}
\DeclareMathOperator{\real}{Re}

\DeclareMathOperator{\Tr}{Tr}

\DeclareMathOperator{\sn}{sn}

\DeclareMathOperator{\vol}{vol}

\newcommand{\ri}{{\rm i}}

\newcommand{\nn}{\nonumber \\}
\newcommand{\eq} {equation}
\newcommand{\eqa} {eqnarray}
\newcommand{\NN} {\nonumber}

\newcommand{\hf}{\frac{1}{2}}

\def\del{\partial}

\def\al{\alpha}

\def\rt#1{\sqrt{#1}}
\def\sitarel#1#2{\mathrel{\mathop{\kern0pt #1}\limits_{#2}}}

\preprint{DESY\ 15-078, HRI/ST/1505}

\title{Large $\bm{N}$ non-perturbative effects in  
$\bm{\mathcal{N}=4}$ superconformal Chern-Simons theories}

\author[a]{Yasuyuki Hatsuda,}
\author[b]{Masazumi Honda}
\author[c]{and Kazumi Okuyama}

\affiliation[a]{DESY Theory Group, DESY Hamburg, \\
Notkestrasse 85, D-22603 Hamburg, Germany}
\affiliation[b]{Harish-Chandra Research Institute, \\
Chhatnag Road, Jhusi, Allahabad 211019, India}
\affiliation[c]{Department of Physics, \\
Shinshu University, Matsumoto 390-8621, Japan}

\emailAdd{yasuyuki.hatsuda@desy.de} 
\emailAdd{masazumihonda@hri.res.in}
\emailAdd{kazumi@azusa.shinshu-u.ac.jp}

\abstract{
We investigate the large $N$ instanton effects of partition functions in a class of $\cN=4$ 
circular quiver Chern-Simons theories on a three-sphere. Our analysis is based on the supersymmetry localization and the Fermi-gas formalism.  
The resulting matrix model can be regarded as a two-parameter deformation of the ABJM matrix model, 
and has richer non-perturbative structures.
Based on a systematic semi-classical analysis, we find analytic expressions of membrane instanton corrections.
We also exactly compute the partition function for various cases
and find some exact forms of worldsheet instanton corrections, 
which appear as quantum mechanical non-perturbative corrections in the Fermi-gas system.
}

\begin{document}

\maketitle
\renewcommand{\thefootnote}{\arabic{footnote}}
\setcounter{footnote}{0}
\setcounter{section}{0}

\section{Introduction}\label{sec:intro}

Over the last couple of years,
many interesting features of non-perturbative effects in M-theory 
on $AdS_4$ background
were discovered via the AdS/CFT correspondence.
Utilizing the powerful techniques of the supersymmetry localization \cite{Pestun} and so-called Fermi gas approach \cite{MP2}
to the partition function of the dual CFT on $S^3$,
now we have a very detailed understanding \cite{HMMO} of 
the non-perturbative effects
in M-theory on $AdS_4\times S^7/\mathbb{Z}_k$,
which is holographically dual to the ABJ(M) theory \cite{ABJM,ABJ}. 

It is realized that the existence of the two types of instanton effects,
worldsheet instantons \cite{Cagnazzo:2009zh} 
and membrane (D2-brane) instantons \cite{DMP2}, 
is necessary for the consistency of the theory. 
From the bulk M-theory perspective,
these two types of instantons are both originating from certain configurations of
M2-branes wrapping some three-cycles, 
but they are distinguished  by the different dependences on the
Planck constant $\hbar =2\pi k$ of the Fermi gas system.
Membrane instantons are already visible 
in the semi-classical small $\hbar$-expansion of the Fermi gas, while
worldsheet instanton effects are non-perturbative in $\hbar$.
One can study such worldsheet instanton effects from the opposite large $\hbar$
regime,
by $1/\hbar$-expansion  of the matrix model
associated with the Fermi gas system, 
which corresponds to the ordinary genus expansion of 
the matrix model with the string coupling given by $g_s=1/\hbar$ \cite{MP1,DMP1,FHM}.
From the viewpoint of this $g_s$-expansion,
the membrane instantons appear as non-perturbative effects in $g_s$.
The pole cancellation mechanism found in \cite{HMO2} is important
for the consistency of these two expansions,
since it guarantees that we can go smoothly from
the small $\hbar$ regime to the large $\hbar$ regime. 
There are also additional contributions, namely the bound states of worldsheet instantons and
membrane instantons \cite{HMO3}, 
which are hard to study from both small $\hbar$ and large $\hbar$ expansions.
Fortunately, in the case of ABJ(M) theory, 
we have a complete understanding of the
worldsheet instantons, membrane instantons, and their bound states,
thanks to the relation to the (refined) topological string on
local $\mathbb{P}^1\times \mathbb{P}^1$ \cite{HMMO,HoO,MaMo,CGM,GHM2}
and exact results for various specific values of the parameters \cite{HMO1,PY,MaMo,HoO}.
(see \cite{Hatsuda:2013yua,Grassi:2013qva} for similar progress in half-BPS Wilson loop)

However, for more general 3d gauge theories with less supersymmetry,
we still do not know detailed structures of the
non-perturbative effects\footnote{%
The only exception so far is the orbifold ABJM theory analyzed in \cite{Honda:2014ica}.
The grand potential of this theory has a simple relation to the one of the ABJM theory.
}.
Some progress along this direction 
has been made in the study of 
an $\mathcal{N}=4$ $U(N)$ gauge theory with $N_f$ fundamental and one adjoint hypermultiplets,
which appears as the worldvolume theory on $N$ D2-branes in the presence of $N_f$ D6-branes.
After applying the localization technique \cite{KWY1,Jafferis,HHL1,Hosomichi},
the partition function of this theory on $S^3$
is reduced to a matrix model, called
the $N_f$ matrix model \cite{GM,MePu}.
In \cite{HO}, using the Fermi gas approach with the identification
$\hbar=N_f$,
the analytic forms of the first few membrane instanton 
corrections of this model were determined.
Worldsheet instantons can also be studied, in principle, by
the genus expansion of the $N_f$ matrix model.
The genus-zero and the genus-one free energies of the $N_f$ matrix model were calculated 
in \cite{GM}, but in practice the computation of the higher 
genus corrections is not so easy.
Instead, the analytic forms of the first few worldsheet instantons 
were found in \cite{HO} from the 
exact computation of the partition functions at finite $N$ up to certain
high $N=N_\text{max}$.
It turned out that the results of the membrane instantons and worldsheet instantons in the $N_f$ matrix model
are quite different from the ABJ(M) case.
In particular, the membrane instanton
coefficients are given by the gamma functions of $N_f$
and quite different from the Gopakumar-Vafa type formula \cite{Gopakumar:1998jq} 
in (refined) topological string, 
where only trigonometric functions of $\hbar$ or $1/\hbar$ appear.
To understand the underlying structure better,
it is desirable to study non-perturbative effects in 
various other models with $\mathcal{N}=4$ supersymmetries.

In this paper we study a special class of $\mathcal{N}=4$ super Chern-Simons matter theories 
in three dimensions \cite{Gaiotto:2008sd,Hosomichi:2008jd}:
a circular quiver gauge theory with the gauge group
$U(N)_k \times U(N)_0^{q-1}\times U(N)_{-k}\times U(N)_0^{p-1}$
and bi-fundamental hypermultiplets one by one between nearest neighboring pairs of gauge groups.
The subscripts in the gauge group 
represent the Chern-Simons level for each  factor.
In this paper, we will refer to this theory as ``$(p,q)$-model''.
It is expected that
the $(p,q)$-model is the low-energy effective theory of $N$ M2-branes probing 
the orbifold $(\mathbb{C}^4 /(\mathbb{Z}_p \times \mathbb{Z}_q))/\mathbb{Z}_k$ \cite{Imamura:2008nn,Imamura:2008dt},
where the orbifolding action
is given by \eqref{eq:orbifold},
and this 
model is dual to M-theory on $AdS_4 \times (S^7  /(\mathbb{Z}_p \times \mathbb{Z}_q))/\mathbb{Z}_k$ 
through the AdS/CFT correspondence.
The $(p,q)$-model can be regarded as a two-parameter deformation of the ABJM theory,
hence it is expected that there is a rich non-perturbative structure in this model.
For instance, the ABJM model corresponds to the $(p,q)=(1,1)$ model,
while the $N_f$ matrix model corresponds\footnote{%
This correspondence is understood from the 3d mirror symmetry \cite{Intriligator:1996ex,Hanany:1996ie}
as explained in \cite{ABJM,Kapustin:2010xq}.
} to the $(p,q)=(1,N_f)$ model with the Chern-Simons level $k=1$. 

We will study the large $N$ instanton effects in the $(p,q)$-model
by analyzing the partition function on $S^3$.
By applying the localization method,
the partition function of the $(p,q)$-model
is reduced to a matrix integral \cite{KWY1,Jafferis,HHL1}, which can be further studied
by the Fermi gas formalism with
the identification $\hbar=2\pi k$. 
Note that the partition function of the $(p,q)$-model
is invariant under the exchange of $p$ and $q$
with fixed $k$.
In the original set-up of the circular quiver gauge theory, $p,q$ and $k$ are all integers,
but at the level of the partition function, we can consider an analytic continuation
of the parameters $(p,q,k)$ to arbitrary real (or complex) numbers. 
The study of the $(p,q)$-model  in the Fermi gas formalism
was initiated in \cite{MN1,MN2}.
The perturbative part of the grand potential was determined in \cite{MN1},
and in \cite{MN2} it was found that
there are three types of membrane instanton corrections
in the grand potential,
\begin{align}
\mathcal{O}(e^{-\frac{2\mu}{p}}),\qquad
\mathcal{O}(e^{-\frac{2\mu}{q}}),\qquad
 \mathcal{O}(e^{-\mu}),\qquad
\label{m2type}
\end{align}
where $\mu$ denotes the chemical potential of the Fermi gas system.
These instantons contribute to the canonical partition function at large $N$
by $\mathcal{O}(e^{-\pi\sqrt{\frac{2qkN}{p}}})$, $\mathcal{O}(e^{-\pi\sqrt{\frac{2pkN}{q}}})$ 
and $\mathcal{O} (e^{-\pi\sqrt{\frac{pqkN}{2}}})$, respectively.
The first two types of instantons are simply related by the exchange of 
$p$ and $q$. In this paper, to study such membrane instanton corrections
we will develop a systematic method for the
$\hbar$-expansion (WKB expansion) of the grand potential. 
Using the data of the WKB expansion, 
we determine the analytic form 
of the leading menbrane instanton correction of the first two types in \eqref{m2type}
for generic $(p,q,k)$, and for the last type in \eqref{m2type} we find the analytic forms
of the leading and the next-to-leading menbrane instanton corrections.
Also, there are worldsheet
instanton corrections of the order
\begin{align}
 \mathcal{O}(e^{-\frac{4\mu}{pqk}}),
\end{align}
which contributes to the canonical partition function by $\mathcal{O} (e^{-2\pi\sqrt{\frac{2N}{pqk}}})$ at large $N$.
From the exact computation of the partition functions at finite $N$,
we find the analytic form of the leading worldsheet instanton correction for generic $(p,q,k)$.
We will see that our results of the menbrane instantons and
worldsheet instantons satisfy the pole cancellation conditions as expected,
and  they also correctly reproduce the known results of the ABJM model and the $N_f$ matrix model
by taking the appropriate limits of the parameters $(p,q,k)$.
Study of the bound state corrections in the $(p,q)$ model is beyond the scope of this paper.

This paper is organized as follows.
In section \ref{sec:Fermi},
after introducing the  Fermi gas formalism of the $(p,q)$-model,
we explain our method of the WKB expansion of the grand potential and 
our algorithm for the exact computation of the partition functions. 
We also comment on the instanton effects
in the $(p,q)$ model from the dual gravity viewpoint. 
Then, we present the results for the $(p,q)=(1,q)$ case in section \ref{sec:1q},
and the results for the general $(p,q)$ case in section \ref{sec:pq}.
In the both cases, we find the analytic forms of the
first few membrane instanton and worldsheet instanton
corrections.
In section \ref{sec:special},
we consider some interesting cases with the special values of $(p,q)$.
We discuss that 
the grand potential for $(p,q)=(1,-1)$ is captured by the refined topological string on a resolved conifold.
We also find an exact closed form expression of the grand
partition function of the $(p,q)=(2,2)$ model at $k=1$. 
Section \ref{sec:con} is devoted to conclusions and discussions.
In appendices \ref{sec:Wigner}
to \ref{sec:variousJ}, we summarize some useful results used in the main text.

\section{Fermi-gas approach to $\cN=4$ quiver CS matrix model}
\label{sec:Fermi}
In this section
we introduce the ideal Fermi gas formalism of the $(p,q)$-model and
discuss how to extract information on the large $N$ non-perturbative effects 
in the corresponding M-theory dual.

\subsection{From matrix model to Fermi-gas}
It is known that
the partition function of the $(p,q)$-model on $S^3$ is reduced to
the matrix integral thanks to 
the supersymmetry localization \cite{KWY1,Jafferis,HHL1},
\begin{\eq}
Z(N,k)  
= \frac{1}{(N!)^{p+q}} \int  \prod_{a=1}^{p+q} \frac{d^N \mu^{(a)}}{(2\pi )^N} 
       \exp{\Bigl[ \frac{ik}{4\pi}\sum_{j=1}^N \left( (\mu_j^{(1)} )^2 -(\mu_j^{(q+1)} )^2 \right)    \Bigr] } 
    \prod_{a=1}^{p+q}
 \frac{\prod_{i<j}\Bigl[ 2\sinh{ \frac{\mu_i^{(a)} -\mu_j^{(a)}}{2}  } \Bigr]^2 }
         {\prod_{i,j}  2\cosh{ \frac{\mu_i^{(a)} -\mu_j^{(a+1)}}{2}  }  }, 
 \label{eq:start}        
\end{\eq}
where $\mu^{(p+q+1)}=\mu^{(1)}$.
This matrix model can be further simplified by using the ideal Fermi gas approach \cite{MP2}.
To be self-contained, here we briefly review the derivation of the Fermi gas representation.
First, by using the Cauchy determinant formula
\[
\frac{\prod_{i<j}\Bigl[ 2\sinh{ \frac{x_i -x_j}{2}  } \Bigr] \Bigl[ 2\sinh{\frac{y_i -y_j}{2} } \Bigr] }
         {\prod_{i,j}  2\cosh{ \frac{x_i -y_j}{2} }  }
  = \sum_\sigma (-1)^\sigma \prod_j 
     \frac{1}{ 2\cosh{\frac{x_j -y_{\sigma (j)}}{2}  }  } ,
\]
we rewrite the partition function as
\begin{\eqa}
Z(N,k)  
&=& \frac{1}{N!} \sum_\sigma (-1)^\sigma \int  \prod_{a=1}^{p+q} \frac{d^N \mu^{(a)}}{(2\pi k)^N} 
       \exp{\Bigl[ \frac{i}{4\pi k}\sum_{j=1}^N \left( (\mu_j^{(1)} )^2 -(\mu_j^{(q+1)} )^2 \right)    \Bigr] } \NN\\
&&\times  \prod_{a=1}^{p+q-1} \Biggl[ \prod_{j=1}^N  \frac{1}{ 2\cosh{ \frac{\mu_j^{(a)} -\mu_j^{(a+1)}}{2k} } } \Biggr] 
\times \prod_{j=1}^N  \frac{1}{ 2\cosh{ \frac{\mu_j^{(p+q)} -\mu_{\sigma (j)}^{(1)}}{2k} } } \NN\\
&=& \frac{1}{N!} \sum_\sigma (-1)^\sigma \int d^N \mu^{(1)} 
        \prod_{j=1}^N \rho (\mu_j^{(1)} ,\mu_{\sigma (j)}^{(1)} ) ,
\end{\eqa}
where we have trivialized most of the permutations and the function $\rho $ is given by
\begin{\eq}
\rho (x,y)
=\frac{1}{2\pi k} \int \prod_{a=2}^{p+q} \frac{d\mu^{(a)}}{2\pi k}
 \frac{e^{\frac{i}{4\pi k} x^2 -\frac{ik}{4\pi}(\mu^{(q+1)} )^2 }}{ 2\cosh{ \frac{x -\mu^{(2)}}{2k} } }
   \prod_{a=2}^{p+q-1} \Biggl[  \frac{1}{ 2\cosh{ \frac{\mu^{(a)} -\mu^{(a+1)}}{2k} } } \Biggr] 
 \frac{1}{ 2\cosh{ \frac{\mu^{(p+q)} -y}{2k} }}  .
\end{\eq}
Thus we can regard the partition function as
an ideal Fermi gas system with the density matrix $\rho (x,y)$.
The expression of $\rho$ is further simplified
if we regard $\rho$ as the matrix element of 
the quantum mechanical operator $\hat{\rho}(\hat{Q},\hat{P})$ 
with $(\hat{Q},\hat{P})$
satisfying the canonical commutation relation
\begin{\eq}
[\hat{Q} ,\hat{P} ] =i\hbar ,\quad \hbar =2\pi k .
\end{\eq}
Then, the density matrix $\rho (x,y)$ is understood as\footnote{%
We are using the following convention
\[
\langle Q| P \rangle =e^{\frac{i}{\hbar}QP},\quad
\int \frac{dQ}{\hbar}|Q\rangle\langle Q| =1 ,\quad
\int \frac{dP}{2\pi}|P\rangle\langle P| =1 .
\]
}
\begin{\eq}
\rho (x,y) =
\frac{1}{\hbar}
\langle x | \hat{\rho}  |y\rangle ,
\end{\eq}
where
\begin{\eq}
\hat{\rho}(\hat{Q},\hat{P})
=
e^{\frac{i}{2\hbar} \hat{Q}^2} \frac{1}{\left( 2\cosh{\frac{\hat{P}}{2}}\right)^q} 
  e^{-\frac{i}{2\hbar} \hat{Q}^2} \frac{1}{\left( 2\cosh{\frac{\hat{P}}{2}}\right)^p} .
\end{\eq}
Using the convenient formula $e^{\frac{i\hat{Q}^2}{2\hbar}} f(\hat{P} )  e^{-\frac{i\hat{Q}^2}{2\hbar}} = f(\hat{P}-\hat{Q})$ and
$e^{\frac{i\hat{P}^2}{2\hbar}} g(\hat{Q} )  e^{-\frac{i\hat{P}^2}{2\hbar}} = g(\hat{Q}+\hat{P})$,
we find
\begin{\eq}
\hat{\rho}(\hat{Q},\hat{P})
=
 e^{-\frac{i}{2\hbar} \hat{P}^2} \frac{1}{\left( 2\cosh{\frac{\hat{Q}}{2}}\right)^q} 
   \frac{1}{\left( 2\cosh{\frac{\hat{P}}{2}}\right)^p} e^{\frac{i}{2\hbar} \hat{P}^2} . 
\end{\eq}
If we perform the similarity transformation
\begin{\eq}
\hat{\rho} \rightarrow 
\Bigl( 2\cosh{\frac{\hat{Q}}{2}}\Bigr)^{q/2}  e^{\frac{i}{2\hbar} \hat{P}^2}  \hat{\rho}
e^{-\frac{i}{2\hbar} \hat{P}^2} \Bigl( 2\cosh{\frac{\hat{Q}}{2}} \Bigr)^{-q/2}  ,
\end{\eq}
then the operator $\hat{\rho}$ is simplified to
\begin{\eq}
\hat{\rho}(\hat{Q},\hat{P})
=
\frac{1}{\left( 2\cosh{\frac{\hat{Q}}{2}}\right)^{q/2}} 
  \frac{1}{\left( 2\cosh{\frac{\hat{P}}{2}}\right)^p}  
\frac{1}{\left( 2\cosh{\frac{\hat{Q}}{2}}\right)^{q/2}} 
\label{eq:rho}  .
\end{\eq}
Note that the partition function is invariant under any similarity transformations.
In the coordinate representation, the density operator \eqref{eq:rho} is expressed by
\be
\rho(x_1,x_2)
=\frac{1}{4\pi^2 k} \frac{1}{\(2\cosh \frac{x_1}{2}\)^{q/2}}
\frac{1}{\(2\cosh \frac{x_2}{2}\)^{q/2}} \B \( \frac{p}{2}+\frac{i(x_1-x_2)}{2\pi k}, \frac{p}{2}-\frac{i(x_1-x_2)}{2\pi k} \),
\label{eq:rho-coord}
\ee
where $\B(x,y)=\Gamma(x)\Gamma(y)/\Gamma(x+y)$ is the Euler beta function.
For $p \in \mathbb{Z}_{>0}$, one can show that this expression reduces to the one\footnote{%
More explicitly, it is given by
\[
\rho(x_1,x_2)
=\left\{ \begin{matrix}
\frac{1}{2(p-1)! \pi k} \frac{1}{\(2\cosh \frac{x_1}{2}\)^{q/2}} \frac{1}{\( 2\cosh \frac{x_2}{2}\)^{q/2}}
\frac{1}{2 \cosh{\frac{x_1 -x_2}{2k}}} 
\prod_{j=1}^{\frac{p-1}{2}} \left[ \left( \frac{x_1 -x_2}{2\pi k}\right)^2 +\frac{(2j-1)^2}{4} \right] &{\rm for\ odd}\ p \cr 
\frac{1}{4(p-1)! \pi^2 k^2} \frac{1}{\(2\cosh \frac{x_1}{2}\)^{q/2}} \frac{1}{\(2\cosh \frac{x_2}{2}\)^{q/2}}
\frac{x_1 -x_2}{2  \sinh{\frac{x_1 -x_2}{2k}}} 
\prod_{j=1}^{\frac{p}{2}-1} \left[ \left( \frac{x_1 -x_2}{2\pi k}\right)^2 +j^2 \right] &{\rm for\ even}\ p \cr 
\end{matrix}\right. .
\]
} in \cite{MN3}.
In particular, for $p=q=1$, it reduces to the density matrix in the ABJM Fermi-gas as expected.
Also, the case for $(p,q,k)=(1,2,1)$ gives
the density matrix for the $U(N)_2 \times U(N+1)_{-2}$ ABJ theory \cite{HoO,Honda,AHS} as explained in \cite{HO}.
Note that the partition function is invariant\footnote{
We can easily show this by the canonical transformation $(\hat{Q}' , \hat{P}' )= (-\hat{P}  , \hat{Q})$ and
the similarity transformation 
$\hat{\rho} \rightarrow ( 2\cosh{\frac{\hat{P}'}{2}} )^{q/2} ( 2\cosh{\frac{\hat{Q}'}{2}} )^{p/2} \hat{\rho} ( 2\cosh{\frac{\hat{P}'}{2}} )^{-q/2} ( 2\cosh{\frac{\hat{Q}'}{2}} )^{-p/2}$.
} under the exchange $p\leftrightarrow q$. This invariance is no longer manifest in the coordinate representation \eqref{eq:rho-coord}.

\subsection{Grand canonical formalism}
\begin{figure}[tb]
\begin{center}
\resizebox{65mm}{!}{\includegraphics{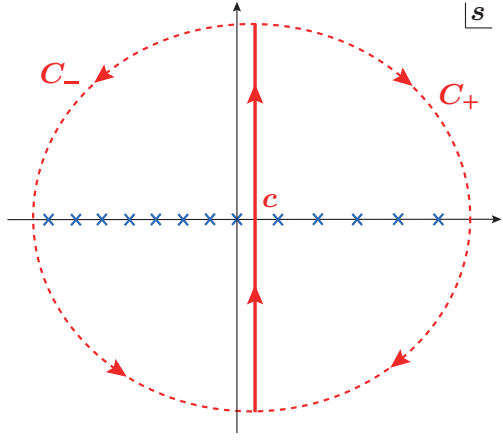}}
\end{center}
\vspace{-0.5cm}
  \caption{One can deform the integration contour in \eqref{eq:J-MB} into a closed path by adding an infinite semi-circle $C_+$ ($C_-$) for $\mu<0$ ($\mu>0)$.
We schematically show the poles of the integrand.}
  \label{fig:contour}
\end{figure}

As was proposed in \cite{MP2}, a useful way to treat this system is to introduce the grand canonical partition function
\be
\Xi(\kappa, k)=1+\sum_{N=1}^\infty \kappa^N Z(N,k) ,\quad \kappa =e^\mu ,
\ee
where $\kappa$ and $\mu$ is the fugacity and the chemical potential, respectively.\footnote{In the following, we will use both $\kappa$ and $\mu$, interchangeably.}
We can return to the canonical partition function by
\begin{\eq}
Z(N,k ) =\frac{1}{2\pi i} \oint \frac{d\kappa}{\kappa^{N+1}}  \Xi(\kappa, k) .
\end{\eq}
The grand partition function is given by the Fredholm determinant
\be
\Xi(\kappa, k)=\det(1+\kappa \hat{\rho})
=\prod_{n=0}^\infty (1+\kappa \lambda_n),
\ee
where $\lambda_n$ ($n\in \mathbb{Z}_{\geq 0}$) are the eigenvalues of $\hat{\rho}$.
The spectral problem for this system is represented by the Fredholm integral equation:
\be
\int_{-\infty}^\infty d x' \, \rho(x, x') \phi_n(x')=\lambda_n \phi_n (x).
\label{eq:eigen-eq}
\ee
It is easy to see that the grand potential is given by
\be
\cJ(\kappa,k)=\log \Xi(\kappa, k)=-\sum_{\ell=1}^\infty \frac{(-\kappa)^\ell}{\ell} \zeta_\rho(\ell),
\label{eq:J}
\ee
where $\zeta_\rho(s)$ is a spectral zeta function defined by
\be
\zeta_\rho(s)=\Tr \hat{\rho}^s=\sum_{n=0}^\infty \lambda_n^s.
\ee
In particular, for $s=\ell \in \mathbb{Z}_{>0}$, it can be computed directly by the multi-integral
\be
\zeta_\rho(\ell)=\mint dx_1\cdots dx_\ell\; \rho(x_1,x_2) \cdots \rho(x_\ell,x_1).
\ee
As in \cite{Hatsuda, MarinoReview}, it is convenient to rewrite \eqref{eq:J} as the following
Mellin-Barnes like integral:
\be
\cJ(\mu,k)=-\int_{c-i \infty}^{c+i \infty} \frac{d s}{2\pi i} \Gamma(s)\Gamma(-s)\zeta_\rho(s) e^{s\mu} .
\label{eq:J-MB}
\ee
The constant $c$ must be taken such that $0<c<1$.
Depending on the sign of $\mu$,
one can deform the integral contour in the following ways.
For $\mu<0$, one can deform the countour by adding an infinite semi-circle $C_+$ as shown in figure~\ref{fig:contour}.
Then the integral can be evaluated by the sum of the residues over all the poles in 
the region $\real s>c$.
As shown in \cite{Hatsuda}, the spectral zeta function $\zeta_\rho(s)$ does not have any poles in 
the region $\real s>0$.
If $0<c<1$, the poles inside the contour are located at $s=\ell$ ($\ell \in \mathbb{Z}_{>0}$),
coming solely from the factor $\Gamma(-s)$, 
and thus we precisely recover the sum \eqref{eq:J}.
On the other hand, for $\mu>0$ one can deform the contour by adding the opposite semi-circle $C_-$ as in figure~\ref{fig:contour}.
In this case, one needs the information of the poles in the region $\real s<c$, in which
$\zeta_\rho(s)$ may have non-trivial poles, and the problem is highly non-trivial for general $k$.
As we will see below, in the semi-classical analysis, we can find all the poles of $\zeta_\rho(s)$
and compute the large $\mu$ expansion systematically.

\subsection{Classical approximation}
Since the ideal Fermi gas formalism provides us with the quantum mechanical description of the
system,
it is useful to consider the semi-classical $\hbar$-expansion, or equivalently the small-$k$ expansion
\be
\cJ_\text{WKB}(\mu,k)=\frac{1}{k} \sum_{n=0}^\infty k^{2n} \cJ^{(n)}(\mu) ,
\label{eq:WKB}
\ee
where $\cJ_\text{WKB}$ is expected to approximate the exact grand potential $\cJ$ 
up to exponentially suppressed corrections in $\hbar$.
Let us start with the classical approximation, namely $\mathcal{O}(k^{-1})$.
Note that the large $\mu$ expansion in this limit has already been analyzed in \cite{MN2}.
Here, we simply re-derive their result by using the Mellin-Barnes integral \eqref{eq:J-MB}. 
In the classical approximation, 
the density operator is given by
\be
\rho_\text{cl}=\frac{1}{\bigl( 2\cosh \frac{Q}{2} \bigr)^q} \frac{1}{\bigl( 2\cosh \frac{P}{2} \bigr)^p}.
\ee
Then, the spectral zeta function can be easily computed by the phase space integral
\be
\zeta_\rho^{(0)}(s)=\int_{-\infty}^\infty \frac{d Q d P}{4\pi^2}
\frac{1}{\bigl( 2\cosh \frac{Q}{2} \bigr)^{qs}} \frac{1}{\bigl( 2\cosh \frac{P}{2} \bigr)^{ps}}
=\frac{1}{4\pi^2} \frac{\Gamma^2(\frac{p s}{2})\Gamma^2(\frac{q s}{2})}{\Gamma(p s) \Gamma(q s)}.
\label{eq:Z0}
\ee
We would like to know the large $\mu$ expansion of the classical grand potential.
Plugging \eqref{eq:Z0} into \eqref{eq:J-MB}, one obtains
\be
\cJ^{(0)}(\mu)=-\frac{1}{4\pi^2} \int_{c-i \infty}^{c+i \infty} \frac{d s}{2\pi i} \Gamma(s)\Gamma(-s)\frac{\Gamma^2(\frac{p s}{2})\Gamma^2(\frac{q s}{2})}{\Gamma(p s) \Gamma(q s)} e^{s\mu}. 
\label{eq:J0}
\ee
If we take the integral contour $\mathcal{C}_-$ in fig.~\ref{fig:contour},
then the leading large $\mu$ contribution comes from the residue at $s=0$:
\be
-\frac{1}{4\pi^2} \Res_{s=0} \Gamma(s)\Gamma(-s)\frac{\Gamma^2(\frac{p s}{2})
\Gamma^2(\frac{q s}{2})}{\Gamma(p s) \Gamma(q s)} e^{s\mu}
=\frac{2}{3\pi^2 pq}\mu^3+\frac{4-p^2-q^2}{6pq}\mu+\frac{p^3+q^3}{\pi^2 p q}\zeta(3) .
\ee
In the region $\real s<0$, the integrand of \eqref{eq:J0} has the following three types of poles:
\be
s=-\frac{2n}{p},\qquad s=-\frac{2n}{q}, \qquad s=-n ,\qquad (n \in \mathbb{Z}_{>0}).
\ee
The residue at $s=-2n/p$ is given by
\be
-\frac{1}{4\pi^2} \Res_{s=-2n/p} \Gamma(s)\Gamma(-s)\frac{\Gamma^2(\frac{p s}{2})
\Gamma^2(\frac{q s}{2})}{\Gamma(p s) \Gamma(q s)} e^{s\mu}
=\frac{1}{2\pi n} \binom{2n}{n} \csc \( \frac{2\pi n}{p} \) 
\frac{\Gamma^2(-\frac{n q}{p})}{\Gamma(-\frac{2n q}{p})}e^{-\frac{2n\mu}{p}}.
\ee
The residue at $s=-2n/q$ is obtained by exchanging $p$ and $q$.
The residue at $s=-n$ is also computed as
\be
-\frac{1}{4\pi^2} \Res_{s=-n} \Gamma(s)\Gamma(-s)\frac{\Gamma^2(\frac{p s}{2})
\Gamma^2(\frac{q s}{2})}{\Gamma(p s) \Gamma(q s)} e^{s\mu}
=\frac{(-1)^{n-1}}{4\pi^2 n} \frac{\Gamma^2(-\frac{n p}{2})}{\Gamma(-n p)}
\frac{\Gamma^2(-\frac{n q}{2})}{\Gamma(-n q)}e^{-n\mu}.
\ee
We conclude that the large $\mu$ expansion of the classical grand potential takes the form
\be
\cJ^{(0)}(\mu)=\frac{2}{3\pi^2 pq}\mu^3+\frac{4-p^2-q^2}{6pq}\mu+\frac{p^3+q^3}{\pi^2 p q}\zeta(3)+\cJ^{(0)}_\text{M2,I}(\mu)
+\cJ^{(0)}_\text{M2,II}(\mu)+\cJ^{(0)}_\text{M2,III}(\mu),
\ee
where 
\be
\ba
\cJ^{(0)}_\text{M2,I}(\mu)&=\sum_{n=1}^\infty \frac{1}{2\pi n} \binom{2n}{n} \csc \( \frac{2\pi n}{p} \) 
\frac{\Gamma^2(-\frac{n q}{p})}{\Gamma(-\frac{2n q}{p})}e^{-\frac{2n\mu}{p}},\\
\cJ^{(0)}_\text{M2,II}(\mu)&=\sum_{n=1}^\infty \frac{1}{2\pi n} \binom{2n}{n} \csc \( \frac{2\pi n}{q} \) 
\frac{\Gamma^2(-\frac{n p}{q})}{\Gamma(-\frac{2n p}{q})}e^{-\frac{2n\mu}{q}},\\
\cJ^{(0)}_\text{M2,III}(\mu)&=\sum_{n=1}^\infty \frac{(-1)^{n-1}}{4\pi^2 n} \frac{\Gamma^2(-\frac{n p}{2})}{\Gamma(-n p)}
\frac{\Gamma^2(-\frac{n q}{2})}{\Gamma(-n q)}e^{-n\mu}.
\ea
\label{eq:J0-M2}
\ee
These expressions reproduce the results in \cite{MN2}.

\subsection{Semi-classical analysis}
Let us proceed to the quantum corrections to the grand potential.
To compute the quantum correction to $\zeta_\rho(s)$, we use the Wigner transform, as in \cite{MP2}.
For a given operator $\hat{A}$, the Wigner transform is defined by
\be
A_\text{W}(Q,P)=\int_{-\infty}^\infty 
 \frac{d Q'}{\hbar}\, 
e^{\frac{i PQ'}{\hbar}} \bra{Q-\frac{Q'}{2}} \hat{A} \ket{Q+\frac{Q'}{2}} .
\ee
The trace of $\hat{A}$ is then given by the phase space integral
\be
\Tr \hat{A}
=\int_{-\infty}^\infty \frac{d Q d P}{2\pi \hbar} A_\text{W}(Q,P).
\ee
Let us apply the Wigner transform to the inverse of the density operator
\be
\hat{\cO}=\hat{\rho}^{-1}=\biggl(2\cosh \frac{\hat{Q}}{2}\biggr)^{q/2}\biggl( 2\cosh \frac{\hat{P}}{2} \biggr)^p \biggl(2\cosh \frac{\hat{Q}}{2}\biggr)^{q/2}.
\ee
As shown in appendix~\ref{sec:Wigner}, the Wigner transform of $\hat{\cO}$ is given by
\be
\cO_\text{W}(Q,P)=\biggl( 4 \cosh^2 \frac{Q}{2}-4 \sin^2 \frac{\pi k \pd_P}{2} \biggr)^{q/2} \biggl( 2 \cosh \frac{P}{2} \biggr)^p.
\label{eq:O_W}
\ee
Using this result, one can easily compute the WKB expansion of $\cO_\text{W}(Q,P)$ up to any order.
We would like to compute the spectral trace
\be
\zeta_\rho(s)=\Tr \hat{\rho}^s=\Tr \hat{\cO}^{-s}.
\ee
Expanding $\hat{\cO}^{-s}$ around $\cO_\text{W}$ as in \cite{MP2,Hatsuda},
the Wigner transform of $\hat{\cO}^{-s}$ is computed by
\be
(\hat{\cO}^{-s})_\text{W}=\sum_{r=0}^\infty \frac{(-1)^r (s)_r}{r!} \cO_\text{W}^{-s-r} [(\hat{\cO}-\cO_\text{W})^r]_\text{W},
\label{eq:f-W-2}
\ee
where $(s)_r$ is the Pochhammer symbol.
To compute the summand in \eqref{eq:f-W-2}, one needs the Wigner transform
of operator products.
The Wigner transform of a product of two operators is computed by
\be
(\hat{A}\cdot \hat{B})_\text{W}=A_\text{W}\star B_\text{W},
\ee
where the Moyal product $\star$ is defined by
\be
\ba
A \star B&:= A(x,p) \exp \left[ \frac{\ri \hbar}{2} ( \stackrel{\leftarrow}{\pd}_x \stackrel{\rightarrow}{\pd}_p
-\stackrel{\leftarrow}{\pd}_p\stackrel{\rightarrow}{\pd}_x  ) \right] B(x,p) \\
&=\sum_{n=0}^\infty \sum_{m=0}^n (-1)^m \binom{n}{m} \frac{1}{n!}\(\frac{\ri \hbar}{2}\)^n
\pd_x^m \pd_p^{n-m} A(x,p) \pd_p^m \pd_x^{n-m} B(x,p) .
\ea
\ee
In this way, one can compute $\Tr \hat{\cO}^{-s}$ up to any desired order, in principle.
However, the integral appearing in \eqref{eq:f-W-2} for general $s$
is complicated and hard to evaluate.
Practically, it is sufficient to compute it for $s \in \mathbb{Z}_{>0}$.
Here, we use an interesting idea in \cite{MN2}.
The quantum correction $\cJ^{(n)} (\mu)$ can be constructed by acting a non-trivial differential operator on
the classical one:
\be
\cJ^{(n)}  (\mu)=\cD^{(n)}  \cJ^{(0)} (\mu), \qquad (n=1,2,\dots),
\label{eq:Jn}
\ee 
where $\cD^{(n)} $ is a differential operator of $\mu$.
Its explicit form up to $n=2$ was computed in \cite{MN2}.
An efficient way to fix this differential operator is as follows.
We first compute the expansion of $\cJ^{(n)}(\kappa)$ around $\kappa=0$.
This can be done by using the formula \eqref{eq:f-W-2} for $s\in \mathbb{Z}_{>0}$.
Taking an ansatz of the form of $\cD^{(n)}$, we try to fix unknown parameters to match the first
several coefficients of $\cJ^{(n)}(\kappa)$.
If the ansatz is correct, the obtained result must reproduce higher order coefficients.
In this way, one can verify the obtained operator up to any desired order.
Using this method, we have indeed fixed the differential operator up to $n=17$.
The result up to $n=4$ is given in appendix~\ref{sec:diff}.

Using this method, we finally find that the semi-classical grand potential has the following large $\mu$ expansion
\be
\cJ_\text{WKB}(\mu,k)
=\cJ_{\rm pert} (\mu ,k)
+\cJ_\text{M2,I}(\mu,k) +\cJ_\text{M2,II}(\mu,k)+\cJ_\text{M2,III}(\mu,k) ,
\label{eq:J-WKB-largemu}
\ee
where $\cJ_{\rm pert}(\mu ,k)$ is the perturbative grand potential  given by
\be
\cJ_{\rm pert} (\mu ,k)
=\frac{C_{p,q}(k)}{3}\mu^3+B_{p,q}(k) \mu+A_{p,q}(k) ,
\label{Jpert}
\ee
with
\be
C_{p,q}(k)=\frac{2}{\pi^2 pq k},\qquad
B_{p,q}(k)=\frac{4-p^2-q^2}{6pq k}+\frac{pq k}{24}.
\ee
The constant part $A_{p,q}(k)$ is a complicated function of $k$,
whose exact form was conjectured in \cite{MN1}
\be
A_{p,q}(k)=\frac{1}{2}\(p^2 A_\text{c}(qk)+q^2 A_\text{c}(pk) \),
\ee
where $A_\text{c}(k)$ is given by \cite{HO} (see also \cite{KEK})
\be
A_\text{c}(k)=\frac{2\zeta(3)}{\pi^2 k}\(1-\frac{k^3}{16}\)
+\frac{k^2}{\pi^2} \int_0^\infty dx \frac{x}{e^{k x}-1}\log(1-e^{-2x}).
\label{eq:A-int}
\ee
There are three types of exponentially suppressed corrections with the following forms
\be
\ba
\cJ_\text{M2,I}(\mu,k)&=\sum_{n=1}^\infty \alpha_n(p,q,k) e^{-\frac{2n\mu}{p}},\qquad
\cJ_\text{M2,II}(\mu,k)=\sum_{n=1}^\infty \alpha_n(q,p,k) e^{-\frac{2n\mu}{q}},\\
\cJ_\text{M2,III}(\mu,k)&=\sum_{n=1}^\infty \beta_n(p,q,k) e^{-n\mu}.
\ea
\label{eq:J-M2-pq}
\ee
Note that $\alpha_n(p,q,k)$ is not symmetric in $p$ and $q$, while $\beta_n(p,q,k)$ is symmetric
(namely, $\alpha_n(p,q,k)\neq \alpha_n(q,p,k)$, $\beta_n(p,q,k)=\beta_n(q,p,k)$).
Our task is to fix these coefficients.

In the later analysis, it is convenient to introduce a function $D(s,p,q,k)$ by
\be
\cD(e^{s\mu})=D(s,p,q,k) e^{s \mu},
\label{eq:D-def}
\ee
where $\cD$ is a generating function of $\cD^{(n)}$,
\be
\cD=1+\sum_{n=1}^\infty k^{2n}\cD^{(n)} .
\ee
The definition \eqref{eq:D-def} means that $D(s,p,q,k)$ is obtained by replacing $\pd_\mu$ in $\cD$ by $s$.
Then, the WKB expansion of the spectral zeta function is simply given by
\be
\zeta_\text{WKB}(s)=\frac{\zeta_\rho^{(0)}(s)}{k}D(s,p,q,k),
\ee
where the classical part $\zeta_\rho^{(0)}(s)$ is given by \eqref{eq:Z0}.
Also, the membrane instanton coefficients in \eqref{eq:J-M2-pq} are generically given by
\be
\alpha_n(p,q,k)=\frac{\alpha_n^{(0)}(p,q)}{k}D\( -\frac{2n}{p} ,p,q,k\), \quad
\beta_n(p,q,k)=\frac{\beta_n^{(0)}(p,q)}{k}D\( -n ,p,q,k\), 
\ee
where $\alpha_n^{(0)}(p,q)$ and $\beta_n^{(0)}(p,q)$ are the classical parts in \eqref{eq:J0-M2}:
\be
\alpha_n^{(0)}(p,q)=\frac{1}{2\pi n} \binom{2n}{n} \csc \( \frac{2\pi n}{p} \) 
\frac{\Gamma^2(-\frac{n q}{p})}{\Gamma(-\frac{2n q}{p})},\quad
\beta_n^{(0)}(p,q)=\frac{(-1)^{n-1}}{4\pi^2 n} \frac{\Gamma^2(-\frac{n p}{2})}{\Gamma(-n p)}
\frac{\Gamma^2(-\frac{n q}{2})}{\Gamma(-n q)}.
\ee
Thus our goal is to determine $D(-2n/p,p,q,k)$ and $D(-n,p,q,k)$.
It is useful to notice the following symmetric property:
\be
D(s,p,q,k)=D(-s,-p,-q,k).
\label{eq:D-sym}
\ee
One can show this by using $\left. \hat{\rho} \right|_{p\rightarrow -p ,q\rightarrow -q}=\hat{\rho}^{-1}$ and
$\zeta_{\rho^{-1}} (-s) =\zeta_\rho (s)$.

\subsection{TBA approach}
There is another approach to compute the grand potential by using the so-called TBA equations.
In \cite{CM}, the semi-classical expansion of the ABJM Fermi-gas was computed in this approach.
In the present situation, for $p=1,2$, we can use this method.
For $p=1$, the density matrix is given by
\be
\left. \rho (x_1,x_2) \right|_{p=1}
= \frac{1}{2\pi k} \frac{e^{-\frac{1}{2}U(x_1)-\frac{1}{2}U(x_2)}}{2\cosh \( \frac{x_1-x_2}{2k} \)},
\quad {\rm with}\ 
U(x)=q \log \left[ 2\cosh \frac{x}{2} \right].
\label{eq:rho-p1}
\ee
The kernel \eqref{eq:rho-p1} has the same form as the one in \cite{Zamolodchikov}, and one can immediately use the result there.
The functional equations are given by
\be
\ba
R_+\( x+\frac{\pi i k}{2} \) R_+\(x- \frac{\pi i k}{2} \) \exp \left[ U\( x+\frac{\pi i k}{2} \)+U\( x-\frac{\pi i k}{2} \) \right]&=1+\eta^2(x),\\
\eta\( x+\frac{\pi i k}{2} \)+\eta\( x-\frac{\pi i k}{2} \)&=-\kappa R_+(x),
\ea
\label{eq:FR-p1-1}
\ee
and
\be
\frac{R_-(x+\frac{\pi i k}{2} )}{R_+ ( x+\frac{\pi i k}{2} )}
-\frac{R_-(x-\frac{\pi i k}{2} )}{R_+ ( x-\frac{\pi i k}{2} )}
=2i k \pd_x \arctan \eta(x).
\label{eq:FR-p1-2}
\ee
The grand potential is computed by
\be
\left. \pd_\kappa \cJ (\kappa,k) \right|_{p=1}
=\frac{1}{4\pi k} \int_{-\infty}^\infty dx (R_+(x)+R_-(x)),
\ee
where an integration constant is fixed by the condition $\cJ(\kappa=0,k)=0$.

For $p=2$, the density matrix is
\be
\left. \rho (x_1,x_2) \right|_{p=2}
=\frac{1}{4\pi^2 k} \frac{1}{\(2\cosh \frac{x_1}{2}\)^{q/2}}
\frac{1}{\(2\cosh \frac{x_2}{2}\)^{q/2}} \frac{\frac{x_1-x_2}{2k}}{ \sinh \frac{x_1-x_2}{2k} }.
\ee
In this case, the density matrix is different from the one in \cite{Zamolodchikov}.
Nevertheless, as explained in appendix~\ref{sec:TBA-p2}, 
we find the following  functional equations
determining the grand potential for $p=2$ in a similar way to the $p=1$ case,
\be
\ba
\xi(x+\pi i k)\xi(x-\pi i k)&=\eta^2(x)-1, \\
\eta(x+\pi i k)+\eta(x-\pi i k)&=2\xi(x) \cosh 2r(x), \\
\frac{w(x+\pi i k)}{\xi(x+\pi i k)}-\frac{w(x-\pi i k)}{\xi(x-\pi i k)}&=2i \pd_x \arccoth \eta(x),
\ea
\label{eq:FR-p2}
\ee
where $\xi(x)$, $\eta(x)$ and $w(x)$ are unknown functions, and $r(x)$ and $t(x)$ are given by
\be
r(x)=\arcsinh \sqrt{\frac{t(x)}{2}},\qquad t(x)=\frac{\kappa}{2(2\cosh\frac{x}{2})^q}.
\label{eq:r}
\ee
The grand potential is then given by
\be
\left. \pd_\kappa \cJ (\kappa,k) \right|_{p=2}
=\frac{1}{\pi \kappa} \int_{-\infty}^\infty dx \, w(x) \sinh^2 r(x).
\label{eq:dJ-p2}
\ee
The functional equations \eqref{eq:FR-p1-1}, \eqref{eq:FR-p1-2} and \eqref{eq:FR-p2} can be solved systematically around $k=0$.
Therefore one can compute the WKB expansion of the grand potential.

\subsection{Non-perturbative corrections: worldsheet instantons}
So far, we have considered the semi-classical analysis, 
which is perturbative in the sense of $\hbar$.
As explained in \cite{MP2,KM}, the grand potential receives quantum mechanical non-perturbative corrections in $k$.
These non-perturbative corrections are caused by the worldsheet instantons in the dual string/M-theory
and invisible in the semi-classical analysis.%
\footnote{Very recently, a new scenario was proposed in \cite{Hatsuda}. This scenario states that the non-perturbative
correction to the grand potential is produced by the perturbative resummation of the spectral zeta function
via the integral transform \eqref{eq:J-MB}.
It would be interesting to explore the worldsheet instanton correction in this approach.} 
In the case of ABJM Fermi-gas, fortunately
these corrections can be  predicted with the help of the topological string on local $\mathbb{P}^1 \times \mathbb{P}^1$.
Interestingly, for some special cases, the Fermi-gas system is related to a quantum mechanical system 
associated with the topological string \cite{GHM1} on certain CY, as will be seen later.
In these cases, it will be possible to predict the worldsheet instanton correction, as in the ABJM Fermi-gas.
However, in general, we do not know such a connection, and
do not have a systematic treatment of these corrections so far.
One approach to compute them is to consider the matrix model computation in the 't Hooft limit, as was performed in \cite{DMP1} for the ABJM matrix model.
In appendix~\ref{app:planar}, we compute the planar free energy of the $(1,q)$ model and
find the worldsheet instanton effect in the planar limit.

Following the argument in \cite{MP2,KM}, one can estimate an order of such a non-perturbative correction.
Let us consider the classical Fermi surface with energy $E$:
\be
H(P,Q)=p \log \( 2 \cosh \frac{P}{2} \)+q \log \( 2\cosh \frac{Q}{2} \)=E.
\label{eq:Fermi-surface}
\ee
This gives an algebraic curve in the phase space.
By rescaling $P=p P' ,\ Q=q Q'$,
this expression becomes
\begin{\eq}
p\log{\left( 2\cosh{\frac{P'}{2p}} \right)} +q\log{\left( 2\cosh{\frac{Q'}{2q}} \right)}=E .
\end{\eq}
In the large $E$ limit, we can approximate the Fermi surface as
\begin{\eq}
H(P',Q') =E \simeq \log{\left( 2\cosh{\frac{P'}{2}} \right)} +\log{\left( 2 \cosh{\frac{Q'}{2}} \right)} ,
\end{\eq}
up to exponentially suppressed correction.
This approximated Hamiltonian leads to the equations of motion
\begin{\eq}
\dot{Q}'= \frac{\del H}{\del P'} = \frac{1}{2} \tanh{\frac{P'}{2}},\quad
\dot{P}'=-\frac{\del H}{\del Q'} = -\frac{1}{2} \tanh{\frac{Q'}{2}} .
\end{\eq}
On the equi-energy orbit $H(P',Q') =E$, $\dot{Q}'$ becomes
\begin{\eq}
\dot{Q}' =\frac{1}{2}\sqrt{1-16\cosh^2 {\frac{Q'}{2}}e^{-2E}} .
\end{\eq}
The solution of this equation of motion is given by
\begin{\eq}
\tanh{\frac{Q'}{2}} = m \sn\left( \frac{t}{4},m\right) ,
\end{\eq}
where $\sn(u,m)$ is the Jacobi's elliptic sine function and $m^2 = 1-16 e^{-2E}$.
As the function of $t$, this has the real period $\omega_1$ and the imaginary period $\omega_2$,
\begin{\eq}
\omega_1 = 16\mathbf{K}(m) ,\quad
\omega_2 = 8i\mathbf{K} (m') ,
\end{\eq}
where $\mathbf{K}(m)$ is the complete elliptic integral of the first kind and $m'^2 =1-m^2$.
Now we consider the complexified Fermi surface, in which we regard $Q'$ and $P'$ as complex variables.
Then, the complexified Fermi surface \eqref{eq:Fermi-surface} determines a Riemann surface,
and we have two kinds of periods associated with this Riemann surface \cite{MP2}.
One of them computes the volume surrounded by the surface \eqref{eq:Fermi-surface}.
We refer to this cycle as the ``B-cycle'' and to the other as the ``A-cycle'', following \cite{KM}.%
\footnote{Note that this convention is opposite to the one in \cite{MP2}.}
The large $E$ behaviors of the periods can be easily estimated.
Noting
\begin{\eq}
P'(t ) 
= -\log{\left( \cosh^2 {\frac{Q'}{2}}e^{-2E} \right)} +\mathcal{O}(e^{-2E}) ,\quad
\omega_1 = 2E  +\mathcal{O}(e^{-2E})  ,
\end{\eq}
the period along the B-cycle is given by
\begin{\eq}
\oint_B P dQ
=\frac{1}{pq} \int_{-\omega_1 /2}^{\omega_1 /2} P'(t)\dot{Q}'(t)dt 
+\mathcal{O}(e^{-\frac{2E}{p}},e^{-\frac{2E}{q}}) 
=\frac{8E^2}{pq} +\mathcal{O}(E) .
\end{\eq}
In order to compute the A-period, it is convenient to use the variables
\begin{\eq}
Q'=i\theta ,\quad t =i\tau .
\end{\eq}
Then we obtain
\begin{\eq}
\oint_A P dQ
=\frac{i}{pq} \int_{-\omega_2 /2i}^{\omega_2 /2i} P'(t)\dot{\theta}(\tau )d\tau 
+\mathcal{O}(e^{-\frac{2E}{p}},e^{-\frac{2E}{q}}) 
=\frac{8\pi i E}{pq} +\mathcal{O}(1) .
\end{\eq}
As explained in \cite{MP2}, the quantum mechanical instanton effect is related to the A-period.
The leading order of this correction is
\be
\exp \left[ \frac{i}{\hbar} \oint_A P dQ \right]=\exp\left[ -\frac{4E}{pqk} \right].
\label{weightmatrix}
\ee
This means that the grand potential receives the non-perturbative correction of order $e^{-\frac{4\mu}{pqk}}$.
We conclude that the worldsheet instanton correction in the present case is expected to take the form
\be
\cJ_\text{WS}(\mu,k)=\sum_{m=1}^\infty d_m(p,q,k)e^{-\frac{4m}{pqk}\mu}.
\label{eq:J-WS}
\ee
We observe that this expectation is precisely consistent with the exact computation of the partition function for various $(p,q,k)$
and the planar solution of the $(1,q)$ model analyzed in appendix~\ref{app:planar}. 
For the very first few coefficients, we can guess the exact forms of $d_m(p,q,k)$, as given in the next two sections.
In addition to the worldsheet instantons, there also exist bound states of the membrane instantons and the worldsheet instantons.
Computation of these bound state contributions is beyond the scope of this work.

\subsection{Exact computation of the partition function}
In this subsection
we present our algorithm for the exact computation of the partition function with fixed integer $p$,
which is a simple generalization of the ABJ(M) case \cite{HMO1,HMO2,HoO}.
First let us recall the formula for the grand partition function 
\begin{\eq}
\Xi (\kappa ,k)
=\exp{\Biggl[ -\sum_{n=1}^\infty \frac{(-\kappa)^n}{n}\Tr \rho^n \Biggr]} ,
\end{\eq}
where we define multiplication and trace of two matrices $\rho_1 ,\rho_2$ as
\begin{\eq}
\rho_1 \rho_2 (x_1 ,x_2 )
=\int_{-\infty}^\infty dy \, \rho (x_1 ,y )\rho_2 (y,x_2 ),\quad
\Tr \rho_1 =\int_{-\infty}^\infty dy \, \rho_1 (y,y ) . 
\end{\eq} 
This formula tells us that
we can exactly compute the canonical partition function with the rank $N$
if we find exact values of $\Tr \rho^n$ with $n=1,\cdots ,N$. 
Therefore, below we explain how to compute the values of $\Tr \rho^n$ exactly
for integer $p$.

\subsubsection{The case of odd $p$}
When $p$ is odd,
the density matrix is given by
\begin{\eq}
\rho(x_1,x_2)
=\frac{1}{2(p-1)! \pi k} \frac{1}{\(2\cosh \frac{x_1}{2}\)^{q/2}} \frac{1}{\(2\cosh \frac{x_2}{2}\)^{q/2}}
\frac{f(x_1 ,x_2 )}{2 \cosh{\frac{x_1 -x_2}{2k}}}  ,
\end{\eq}
where
\begin{\eq}
f(x_1 ,x_2 )
= \prod_{j=1}^{\frac{p-1}{2}} \left[ \left( \frac{x_1 -x_2}{2\pi k}\right)^2 +\frac{(2j-1)^2}{4} \right]
=\sum_{j,j'=0}^{p-1} f_{j j'}x_1^j x_2^{j'} .
\end{\eq}
Then, we rewrite the density matrix as
\begin{\eq}
\rho (x_1 ,x_2 )
=\frac{E(x_1 )E(x_2 )}{M(x_1 ) +M(x_2 )} \sum_{j,j' =0}^{p-1} f_{j j'}x_1^j x_2^{j'} ,
\end{\eq}
where
\begin{\eq}
E(x) = \frac{e^{\frac{x}{2k}}}{\left( 2\cosh{\frac{x}{2}}\right)^{q/2}} ,\quad
M(x) = 2(p-1)! \pi k e^{\frac{x}{k}} .
\end{\eq}
This relation is also schematically represented by
\begin{\eq}
\{ M ,\rho \} =\sum_{j,j'=0}^{p-1} f_{j j'}  ( x^j E ) \otimes ( x^{j'} E ) .
\end{\eq}
Here we regard $\rho ,M$ and $x^j E $ as
the symmetric matrix, diagonal matrix and vector, respectively,
whose indices are the coordinates $(x_1 ,x_2 )$.
Applying this relation iteratively, we find
\begin{\eq}
\{ M ,\rho^n ] =\sum_{\ell =0}^{n-1}(-1)^\ell 
\sum_{j,j'=0}^{p-1}f_{j j'} ( \rho^\ell x^j E )\otimes (\rho^{n-1-\ell}x^{j'}E ) ,
\end{\eq}
which is equivalent to
\begin{\eq}
\rho^n (x_1 ,x_2 )
= \frac{1}{M(x_1 ) +(-1)^{n-1} M(x_2 )} 
\sum_{\ell =0}^{n-1} (-1)^\ell E(x_1) E(x_2)
\sum_{j,j'=0}^{p-1}f_{j j'}  \phi_\ell^{(j)}(x_1 ) \phi_{n-1-\ell}^{(j')}(x_2 ) .
\end{\eq}
Here $\phi_\ell^{(j)}(x)$ satisfies the recursion relation
\begin{\eqa}
\phi_{\ell}^{(j)}(x)
&=&\frac{1}{E(x)} \int_{-\infty}^\infty \frac{dy}{2\pi k}  \rho (x,y) E(y) \phi_{\ell -1}^{(j)}(y) ,
\label{eq:recursion}
\end{\eqa}
with the initial condition
\begin{\eq}
\phi_{0}^{(j)} = x^j . 
\label{eq:initial}
\end{\eq}
Once we know the series of functions $\phi_{\ell}^{(j)}(x)$,
we can compute $\Tr\rho^n$ systematically.

\subsubsection{The case of even $p$}
For even $p$, the density matrix is given by 
\begin{\eq}
\rho(x_1,x_2)
=\frac{1}{4(p-1)! \pi^2 k^2} \frac{1}{\(2\cosh \frac{x_1}{2}\)^{q/2}} \frac{1}{\(2\cosh \frac{x_2}{2}\)^{q/2}}
\frac{f(x_1 ,x_2 )}{2 \sinh{\frac{x_1 -x_2}{2k}}}  ,
\end{\eq}
where
\begin{\eq}
f(x_1 ,x_2 )
=(x_1 -x_2 ) \prod_{j=1}^{\frac{p}{2}-1} \left[ \left( \frac{x_1 -x_2}{2\pi k}\right)^2 +j^2 \right] 
=\sum_{j,j'=0}^{p-1} f_{j j'}x_1^j x_2^{j'} .
\end{\eq}
Then, we rewrite the density matrix as
\begin{\eq}
\rho (x_1 ,x_2 )
=\frac{E(x_1 )E(x_2 )}{M(x_1 ) -M(x_2 )} \sum_{j,j' =0}^{p-1} f_{j j'}x_1^j x_2^{j'} ,
\end{\eq}
where
\begin{\eq}
E(x) = \frac{e^{\frac{x}{2k}}}{\left( 2\cosh{\frac{x}{2}}\right)^{q/2}} ,\quad
M(x) = 4(p-1)! \pi^2 k^2 e^{\frac{x}{k}} .
\end{\eq}
This relation has the similar structure as in the odd $p$ case:
\begin{\eq}
[ M ,\rho ] =\sum_{j,j'=0}^{p-1} f_{j j'}  ( x^j E ) \otimes ( x^{j'} E ) .
\end{\eq}
Hence, a similar argument leads us to
\begin{\eq}
\rho^n (x_1 ,x_2 )
= \frac{1}{M(x_1 ) - M(x_2 )} 
\sum_{\ell =0}^{n-1} (-1)^\ell E(x_1) E(x_2)
\sum_{j,j'=0}^{p-1}f_{j j'}  \phi_\ell^{(j)}(x_1 ) \phi_{n-1-\ell}^{(j')}(x_2 ) ,
\label{eq:rho-power-even}
\end{\eq}
where $\phi_\ell^{(j)}(x)$ satisfies formally the same relations \eqref{eq:recursion} and \eqref{eq:initial} as in the odd $p$ case.

\subsubsection{Free energy for various $(p,q,k)$}
Using the method explained in the previous subsections,
we have computed the exact values of partition functions
up to certain $N=N_\text{max}$, for various $(p,q,k)$.
For example, for the case of $(p,q,k)=(1,2,2), (2,3,1), (2,3,2)$, 
we have computed the exact partition functions
up to $N_\text{max}=66, 32,33$, respectively.
Some examples of the exact values of partition functions can be found
in appendix \ref{sec:variousJ}.
These exact data are very useful to extract the instanton corrections as in \cite{HMO2}.

\begin{figure}[tb]
\begin{center}
\resizebox{100mm}{!}{\includegraphics{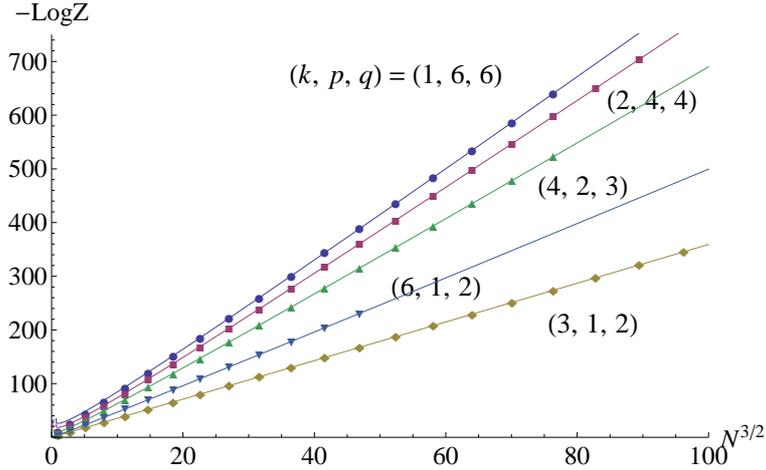}}
\end{center}
\vspace{-0.5cm}
  \caption{The free energy $(-\log{Z})$ is plotted against $N^{3/2}$ for some $(k,p,q)$. 
The straight lines show the perturbative free energy $(-\log{ Z_\text{pert}})$.}
  \label{fig:pert}
\end{figure}

From the general argument in the Fermi gas approach \cite{MP2}, 
in the large $N$ limit
the partition function $Z(N,k)$ behaves as
\begin{align}
 Z(N,k)= Z_\text{pert}(N,k)\cdot\big[1+\mathcal{O}(e^{-\rt{N}})\big],
\end{align}
where the perturbative part is given by the Airy function
\cite{FHM,MP2}
\begin{align}
 Z_\text{pert}(N,k)=C_{p,q}(k)^{-1/3}e^{A_{p,q}(k)}\text{Ai}\Big[C_{p,q}(k)^{-1/3}(N-B_{p,q}(k))\Big].
\label{eq:pert}
\end{align}
The constants $A_{p,q}(k)$, $B_{p,q}(k)$ and $C_{p,q}(k)$ appearing in $Z_\text{pert}(N,k)$ are none other than the coefficients
of the perturbative part of the grand potential \eqref{Jpert}.
In fig.~\ref{fig:pert},
we plot the exact values of the free energy $(-\log{Z})$ for some $(k,p,q)$
with the perturbative free energy $(-\log{ Z_\text{pert}})$.
We can easily see that
the exact free energy shows a good agreement with the perturbative free energy
since their difference is exponentially suppressed in the large $N$ regime.
We also observe that the free energy scales like $N^{3/2}$ for large $N$ as expected 
from the AdS/CFT correspondence as found earlier in \cite{HKPT,Jafferis:2011zi,MP2}.
The perturbative free energy also contains
the log-correction $\frac{1}{4}\log{N}$ in subsubleading large $N$ correction
as expected from the one-loop analysis on the gravity side \cite{Bhattacharyya:2012ye}.

\subsection{A comment on the gravity dual}
The $(p,q)$-model is expected to be the effective theory of $N$ M2-branes 
on the orbifold ($\mathbb{C}^4 /(\mathbb{Z}_{p}\times \mathbb{Z}_q ) )/\mathbb{Z}_k$ \cite{Imamura:2008nn}:
\begin{\eqa}
&&\phi_A :(z_1 ,z_2 ,z_3 , z_4) \sim (e^{\frac{2\pi i}{q}}z_1 ,e^{-\frac{2\pi i}{q}} z_2, z_3 ,z_4 ) ,\NN\\
&&\phi_B :(z_1 ,z_2 ,z_3 , z_4) \sim (z_1 , z_2, e^{\frac{2\pi i}{p}}z_3 , e^{-\frac{2\pi i}{p}}z_4 ) ,\NN\\
&&\phi_C : (z_1 ,z_2 ,z_3 , z_4) \sim 
(e^{\frac{2\pi i}{kq}}z_1 ,e^{-\frac{2\pi i}{kq}} z_2, e^{\frac{2\pi i}{kp}} z_3 ,e^{-\frac{2\pi i}{kp}}z_4 ) .
\label{eq:orbifold}
\end{\eqa}
This implies that
the $(p,q)$-model is dual to M-theory on $AdS_4 \times (S^7 /(\mathbb{Z}_{p}\times \mathbb{Z}_q ) )/\mathbb{Z}_k$
with the metric
\begin{\eq}
ds^2 = \frac{R^2}{4} ds_{AdS_4}^2 +R^2 ds^2_{(S^7 /(\mathbb{Z}_{p}\times \mathbb{Z}_q ) )/\mathbb{Z}_k} ,
\end{\eq}
where
\begin{\eq}
R = (32\pi^2 kpq N)^{1/6} l_p .
\end{\eq}

Since this background has many nontrivial 3-cycles,
we could have discrete holonomies of the 3-form potential along the cycles as in the ABJ theory \cite{ABJ}.
For Imamura-Kimura type theory with equal ranks of gauge groups (without fractional branes),
the discrete holonomies depend on the ordering of 5-branes in its type IIB brane construction.
This has been studied in detail in \cite{Imamura:2009ur} by analyzing monopole operators in general Imamura-Kimura type theory.
According to the formula in \cite{Imamura:2009ur},
we expect that the gravity dual of the $(p,q)$-model does not have the discrete holonomies.

There are some predictions on the free energy $-\log{Z}$ from the gravity side.
First the free energy of the classical SUGRA with the boundary $S^3$ 
is given by (see e.g. \cite{Marino:2011nm})
\begin{\eq}
F_{\rm SUGRA}
= \frac{\pi\sqrt{2kpq}}{3} N^{3/2}.
\end{\eq}
Also by one-loop analysis of the 11d SUGRA on $AdS_4 \times X_7$ with the smooth 7d manifold $X_7$,
it is known that the one-loop free energy contains the following universal log-correction\footnote{
When $X_7$ has fixed points as in our case, there might be extra massless degrees of freedom and
the logarithmic behavior could change.
However, the agreement to the CFT side implies absence of such extra contributions.
} \cite{Bhattacharyya:2012ye}
\begin{\eq}
\frac{1}{4}\log{N} .
\end{\eq}
On the CFT side,
this behavior comes from the the Airy functional behavior \eqref{eq:pert} in the perturbative free energy.

Next we give some comments on nonperturbative effects.
Let us first recall the ABJM case. 
For the ABJM case ($p=q=1$),
if we identify the M-theory circle with the orbifolding direction by $\mathbb{Z}_k$ 
and shrink the circle,
then the 11d supergravity on $AdS_4 \times S^7 /\mathbb{Z}_k$ becomes
the type IIA supergravity on $AdS_4 \times \mathbb{CP}^3$.
In the type II superstring on $AdS_4 \times \mathbb{CP}^3$,
we have 
worldsheet instanton effect, which comes from fundamental string wrapping 
the nontrivial 2-cycle $\mathbb{CP}^1$ in $\mathbb{CP}^3$ \cite{Cagnazzo:2009zh}.
From the M-theory viewpoint,
this corresponds to an M2-brane wrapping the non-trivial 3-cycle $S^3 /\mathbb{Z}_k$
in $S^7 /\mathbb{Z}_k$.
For the general $(p,q)$ case,
we also expect that there are similar non-perturbative effects
as in the ABJM case.
Note that the orbifold \eqref{eq:orbifold} includes
the nontrivial 3-cycle $(S^3 /(\mathbb{Z}_{p}\times \mathbb{Z}_q)/\mathbb{Z}_k)$,
which is obtained by taking $z_2 =z_4  =0$, for example.
This implies the presence of non-perturbative effect
coming from M2-brane wrapping this cycle,
whose weight is given by\footnote{
Note that the tension $T_{\rm M2}$ of the M2-brane is given by $T_{M2}=1/(4\pi^2 l_p^3 )$.
} 
\begin{\eq}
\exp{\Biggl[ -T_{\rm M2}{\rm Vol}\left( (S^3 /(\mathbb{Z}_{p}\times \mathbb{Z}_q ) )/\mathbb{Z}_k \right) \Biggr]}
=\exp{\Biggl[ -\frac{R^3}{2kpq} \Biggr]}
=\exp{\Biggl[ -2\pi \sqrt{\frac{2N}{kpq}} \Biggr]} .
\label{sugraweight}
\end{\eq}
Since the 3-cycle becomes two-dimensional in the large-$k$ limit,
we expect that this effect corresponds to the worldsheet instanton effect
described by the fundamental string wrapping a 2-cycle in the type IIA superstring theory.
One can see that the weight \eqref{sugraweight}
of the worldsheet instanton effect
computed from the gravity side correctly reproduces
the weight \eqref{weightmatrix}
obtained by the matrix model, after changing the variable from the canonical to the grand
canonical ensemble.
This is also consistent with the planar solution of the $(1,q)$ model computed in appendix \ref{app:planar}.

\section{Results on the $(1,q)$-model}
\label{sec:1q}
In this section, we summarize some explicit results in the case of $p=1$.
In this case, the system can be thought of as  a one-parameter deformation of the 
ABJM Fermi-gas by $q$,
or the deformation of the $N_f$ matrix model \cite{MePu, GM} by $k$.
Therefore, the results for $k=1$ (and $q=N_f$) must reproduce those in the $N_f$ matrix model.
Similarly, in the limit $q \to 1$ (with general $k$), the results must also reproduce those in the ABJM Fermi-gas.

\subsection{Membrane instanton corrections}
Let us first consider the membrane instanton corrections.
First of all, one notices that the classical membrane instanton corrections $\cJ^{(0)}_\text{M2,I}$ and $\cJ^{(0)}_\text{M2,III}$
in \eqref{eq:J0-M2} are divergent in the limit $p \to 1$.
As was shown in \cite{MN2}, these divergences are, however, canceled by each other.
One finds that after the cancellation, the finite part is given by 
\be
\widehat{\cJ}^{(0)}_\text{M2,I}(\mu)=\lim_{p \to 1} (\cJ^{(0)}_\text{M2,I}+\cJ^{(0)}_\text{M2,III})
=\sum_{n=1}^\infty (\gamma^{(0)}_n(q) \mu+\delta^{(0)}_n(q))e^{-2n \mu},
\ee
where
\be
\ba
\gamma_n^{(0)}(q)&=-\frac{1}{2\pi^2 n} \binom{2n}{n} \frac{\Gamma^2(-n q)}{\Gamma(-2n q)},\\
\delta_n^{(0)}(q)&=-\frac{1}{4\pi^2 n^2} \binom{2n}{n} \frac{\Gamma^2(-n q)}{\Gamma(-2n q)}
\bigl[ 1+2n(H_n-H_{2n})+2nq \bigl(\psi(-n q)-\psi(-2nq) \bigr) \bigr].
\ea
\ee
Here $H_n$ is the $n$-th harmonic number, and $\psi(z)=\pd_z \log \Gamma(z)$ is the digamma function.
There is no limit problem for $\cJ^{(0)}_\text{M2,II}$:
\be
\ba
\widehat{\cJ}^{(0)}_\text{M2,II}(\mu)&=\lim_{p \to 1} \cJ^{(0)}_\text{M2,II}=\sum_{n=1}^\infty \alpha_n^{(0)}(q) e^{-\frac{2n\mu}{q}},\\
\alpha_n^{(0)}(q)&=\frac{1}{2\pi n} \binom{2n}{n} \csc \( \frac{2\pi n}{q} \) 
\frac{\Gamma^2(-\frac{n}{q})}{\Gamma(-\frac{2n}{q})}.
\ea
\ee
Therefore, in the case of $p=1$, the large $\mu$ expansion \eqref{eq:J-WKB-largemu} reduces to
\be
\cJ_\text{WKB}^{p=1}(\mu,k)=\frac{C_{1,q}(k)}{3}\mu^3+B_{1,q}(k) \mu+A_{1,q}(k)+\widehat{\cJ}_\text{M2,I}(\mu,k)
+\widehat{\cJ}_\text{M2,II}(\mu,k) ,
\ee
where
\be
\ba
\widehat{\cJ}_\text{M2,I}(\mu,k)&=\sum_{n=1}^\infty (\gamma_n(q,k)\mu+\delta_n(q,k))e^{-2n\mu},\\
\widehat{\cJ}_\text{M2,II}(\mu,k)&=\sum_{n=1}^\infty \alpha_n(q,k) e^{-\frac{2n \mu}{q}}.
\ea
\ee
Note that $\alpha_n(q,k)=\alpha_n(q,1,k)$, not $\alpha_n(1,q,k)$.
Acting the differential operator $\cD^{(n)}$ in appendix~\ref{sec:diff} on the classical grand potential, one can find the WKB expansion of each coefficient.

\paragraph{Coefficients of $e^{-2n\mu}$.} 
To find the WKB expansions of $\gamma_n(q,k)$ and $\delta_n(q,k)$, we need to compute
\be
\cD^{(n)} (\mu e^{-2n \mu}) \quad \text{and} \quad \cD^{(n)} (e^{-2n \mu}).
\label{eq:D-act}
\ee 
The computation of $\gamma_n(q,k)$ is relatively easy, compared to 
the other instanton coefficients $\al_n(q,k),\delta_n(q,k)$.
It takes the form
\be
\gamma_n(q,k)=\frac{\gamma_n^{(0)}(q)}{k} D(-2n,1,q,k),
\label{gammadef}
\ee
where, as mentioned in the previous section, $D(-2n,1,q,k)$ is obtained by replacing $\pd_\mu$ in the differential operator $\cD$ by $-2n$.
Therefore its WKB expansion is
\be
\ba
D(-2n,1,q,k)&=
1-\frac{n^2 (1+2 n) q^2}{24 (-1+2 n q)}(\pi k)^2\\
&\quad+\frac{n^3 (1+2 n) q^3 \left(4-24 n+4 q-3 n q+14 n^2 q\right)}{5760 (-3+2 n q) (-1+2 n q)}(\pi k)^4
+\cO(k^{6}) .
\ea
\ee
Since we have fixed $\cD^{(n)}$ up to $n=17$, we can compute the WKB expansion
up to $\cO(k^{34})$.
All of the following results reproduce the correct WKB expansions up to this order.
From the WKB data, we find analytic expressions for $n=1,2$:
\be
\ba
D(-2,1,q,k)&={}_2F_1\( -\frac{q}{2}, -\frac{q}{2}; \frac{1}{2}-q; \sin^2 \frac{\pi k}{2} \), \\
D(-4,1,q,k)&= \frac{1}{3}\,{}_2F_1\( -q, -q; \frac{1}{2}-2q; \sin^2 \frac{\pi k}{2} \) 
+\frac{2}{3}\,{}_2F_1\( -q, -2q; \frac{1}{2}-2q; \sin^2 \frac{\pi k}{2} \).
\label{D1qk}
\ea
\ee
One can check that,
in the limit $k\to 1$, $\gamma_1(q,k)$ and $\gamma_2(q,k)$ 
given by \eqref{gammadef} with \eqref{D1qk}
reduce to
\be
\ba
\gamma_1(q,1)=-\frac{1}{2\pi^2} \frac{\Gamma^2(-q/2)}{\Gamma(-q)},\qquad
\gamma_2(q,1)=-\frac{1}{4\pi^2}\(1+\frac{2}{\cos \pi q}\) \frac{\Gamma^2(-q)}{\Gamma(-2q)}, 
\ea
\ee
which are in perfect agreement with the known results of $N_f$-matrix model in \cite{HO}.

The constant part $\delta_n(q,k)$ is more involved.
Both of \eqref{eq:D-act} contribute to $\delta_n(q,k)$.
The latter contribution is just $D(-2n,1,q,k)$.
A simple computation shows that the $\mu$-independent contribution of the former is given by $\pd_s D(s,1,q,k)|_{s=-2n}$.
We conclude that 
\be
\delta_n(q,k)
=\frac{\delta_n^{(0)}(q)}{k}\Bigl( D(-2n,1,q,k)+\pd_s D(s,1,q,k)|_{s=-2n} \Bigr) ,
\ee
where the WKB expansion of $\pd_s D(s,1,q,k)|_{s=-2n}$ is given by
\be
\ba
&\pd_s D(s,1,q,k)|_{s=-2n}=\frac{n q^2 \left(-1-3 n+n q+4 n^2 q\right)}{24 (-1+2 n q)^2}(\pi k)^2\\
&\qquad-\frac{n^2 q^3}{2880 (-3+2 n q)^2 (-1+2 n q)^2}\bigl[ -3 (-3 + 16 n + 60 n^2) +(9-n+126 n^2 \\
&\qquad\quad+510 n^3)q-2 n (8+13 n+48 n^2+212 n^3)q^2+2 n^2 (2 + 5 n  \\
&\qquad\quad+ 12 n^2 + 56 n^3)q^3 \bigr] (\pi k)^4 +\cO(k^6).
\ea
\ee
It is difficult to guess an exact form of this expansion even for $n=1$.

\begin{figure}[tb]
\begin{center}
\begin{tabular}{cc}
\resizebox{65mm}{!}{\includegraphics{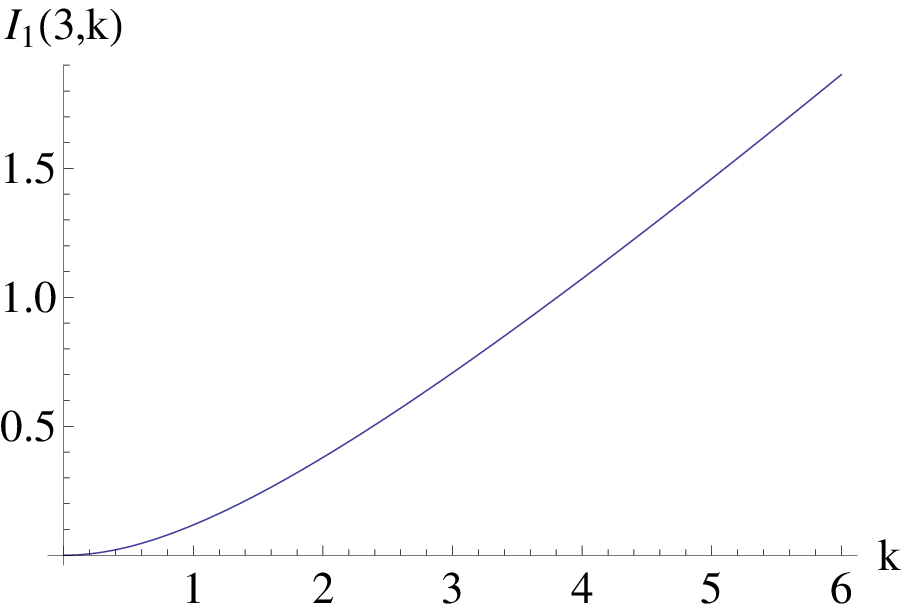}}
\hspace{4mm}
&
\resizebox{72mm}{!}{\includegraphics{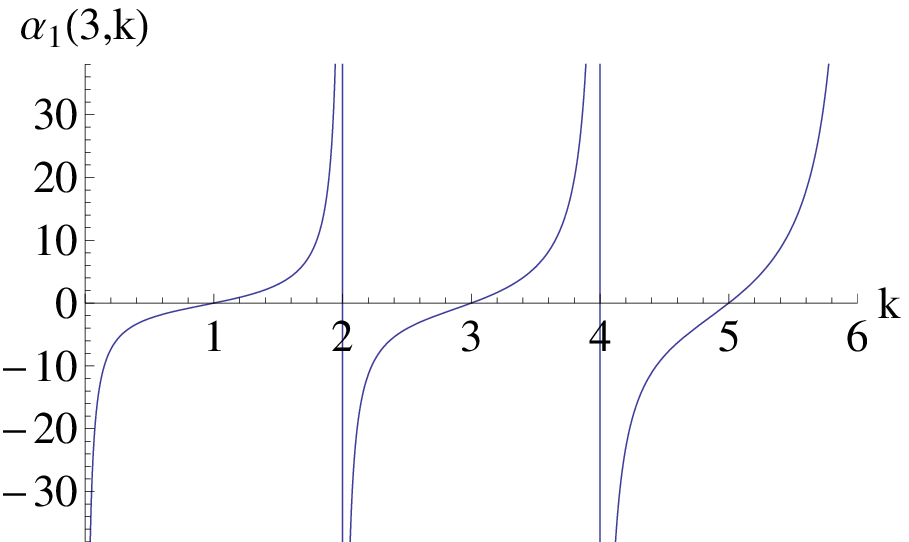}}
\vspace{-5mm}
\end{tabular}
\end{center}
  \caption{We plot $I_1(3,k)$ (Left) and $\alpha_1(3,k)$ (Right) as functions of $k$. In general, $\alpha_1(q,k)$ has simple poles at even $k$.}
  \label{fig:alpha1-q3}
\end{figure}

\paragraph{Coefficients of $e^{-2n\mu /q}$.} 
Next, let us consider the second type correction $\widehat{\cJ}_\text{M2,II}$.
The coefficient $\alpha_n(q,k)$ is given by
\be
\alpha_n(q,k)=\frac{\alpha_n^{(0)}(q)}{k}D\(-\frac{2n}{q},1,q,k\). 
\ee
The WKB expansion of $D(-2n/q,1,q,k)$ is given by
\be
\ba
D\(-\frac{2n}{q},1,q,k\)&= 1-\frac{n^2  (2 n+q)}{24 q (-1+2 n)}(\pi k)^2 \\
&\quad+\frac{n^3 (2 n+q) \left(-24 n+14 n^2+4 q-3 n q+4 q^2\right)}{5760 q^2 (-3+2 n) (-1+2 n)}(\pi k)^4+\cO(k^6).
\ea
\ee
It is not easy to find out an exact expression of this expansion, but for $n=1$ we find a surprisingly simple expression
in terms of the $q$-gamma function,
\be
D\(-\frac{2}{q},1,q,k\)=\frac{\Gamma^2(1+1/q)}{\Gamma(1+2/q)}\frac{\Gamma_{\mathfrak{q}}(1+2/q)}{\Gamma_{\mathfrak{q}}^2(1+1/q)}\mathfrak{q}^{-\frac{1}{2q^2}},
\quad \mathfrak{q}=e^{\pi i kq},
\label{r1qGamma}
\ee
where $\Gamma_{\mathfrak{q}}(z)$ is the $q$-gamma function defined in appendix~\ref{sec:qGamma}.
 As we will see in the next section,
this conjecture comes from the 
analysis of this instanton coefficient 
for the $1/q\in \mathbb{Z}$ case
\eqref{qfactorial}.
When $kq \in \mathbb{R}$, there is a subtlety of the definition of the $q$-gamma function due to $|\mathfrak{q}|=1$. As explained
in appendix~\ref{sec:qGamma}, one can define
the $q$-gamma function with $|\mathfrak{q}|=1$
by regularizing the infinite product using the zeta-function regulariation.
Using the result \eqref{ourqgamma}
in appendix~\ref{sec:qGamma}, one obtains the all-order WKB expansion
\be
\ba
&D\(-\frac{2}{q},1,q,k\)=\exp \left[\sum_{n=1}^\infty\frac{ (-1)^{n-1}B_{2n}}{2n(2n+1)!}
\(2B_{2n+1}\(1+q^{-1}\)-B_{2n+1}\(1+2q^{-1}\) \) (\pi k q)^{2n} \right] \\
&\qquad=\frac{\pi k}{2} \cot  \( \frac{\pi k}{2}\) 
\exp \left[\sum_{n=1}^\infty\frac{ (-1)^{n-1}B_{2n}}{2n(2n+1)!}
\(2B_{2n+1}\(q^{-1}\)-B_{2n+1}\(2q^{-1}\) \) (\pi k q)^{2n} \right] ,
\label{eq:alpha1-WKB}
\ea
\ee
where $B_n$ and $B_n(x)$ are the Bernoulli number and the Bernoulli polynomial, respectively.
From the first line to the second line in \eqref{eq:alpha1-WKB}, we have used the identity
\be
B_{m}(1+w)=B_{m}(w)+m w^{m-1}.
\label{bernouli-shift}
\ee
By using an integral representation of the Bernoulli polynomial \cite{Bernoulli}
\be
B_{2n+1}(w)=(-1)^{n-1}(2n+1) \int_0^\infty dt\, \frac{\sin 2\pi w}{\cosh 2\pi t-\cos 2\pi w}t^{2n},\qquad 0 \leq w \leq1,
\label{eq:Bernoulli-int}
\ee
one can perform the resummation of \eqref{eq:alpha1-WKB}.
For $q \geq 2$, one easily finds
\be
D\(-\frac{2}{q},1,q,k\)=\frac{\pi k}{2} \cot  \( \frac{\pi k}{2}\) \exp [ I_1(q,k)], \qquad q\geq 2,
\label{eq:alpha1-int}
\ee
where
\be
I_1(q,k)=\int_0^\infty dt \frac{4\sin^2 \frac{\pi}{q} \sin \frac{2\pi}{q} (\cosh 2\pi t+2\cos^2 \frac{\pi}{q} )}
{(\cosh 2\pi t-\cos \frac{2\pi}{q} )(\cosh 2\pi t-\cos \frac{4\pi}{q}) }
\log \( \frac{\sinh(\pi k q t/2)}{\pi k q t/2} \).
\ee
For $1\leq q \leq 2$, to use the integral representation \eqref{eq:Bernoulli-int}, we further need to shift the argument 
by using \eqref{bernouli-shift} with $w=2q^{-1}-1$.
Then we find
\be
D\(-\frac{2}{q},1,q,k\)=\frac{\pi k}{2} \cot  \( \frac{\pi k}{2}\) \frac{\sin \frac{\pi k}{2}(2-q)}{\frac{\pi k}{2} (2-q)} 
\exp [ I_1(q,k)], \qquad 1\leq q \leq 2,
\label{eq:alpha1-int2}
\ee
where $I_1(q,k)$ is the same as above.
In a similar way, one also obtains the integral representation for $0<q<1$.
These integral representations are very useful to understand the pole structure of $\alpha_1(q,k)$.
Since the integrand of $I_1(q,k)$ is exponentially dumped for large $t$
and does not have singularities in the integral domain for $q\neq 1,2$ and finite $k$,
$I_1(q,k)$ takes in a finite real value and hence does not have any singularities for $k>0$.
Thus the singularities of $\alpha_1(q,k)$ come only from the cotangent factor in \eqref{eq:alpha1-int} or \eqref{eq:alpha1-int2}.
In figure~\ref{fig:alpha1-q3}, we illustrate these for $q=3$.

\subsection{Worldsheet instanton corrections}
In the case of $p=1$, the worldsheet instanton corrections take the form
\be
\left. \cJ_\text{WS}(\mu,k) \right|_{p=1}
=\sum_{m=1}^\infty d_m(1,q,k)e^{-\frac{4m}{qk}\mu}.
\ee
From a consistency with the results for various integral $(q,k)$ (see appendix~\ref{sec:variousJ}), we conjecture the exact form of
 worldsheet instanton coefficients for $m=1,2$:
\be
\ba
d_1(1,q,k)&=\frac{q}{\sin \frac{2\pi}{k} \sin \frac{2\pi}{q k} }, \\
d_2(1,q,k)&=-\frac{1}{2\sin^2\frac{2\pi}{qk}}-\frac{q^2}{\sin^2\frac{2\pi}{k}}
+\frac{q\sin\frac{6\pi}{qk}}{2\sin\frac{4\pi}{k}\sin\frac{2\pi}{qk}\sin\frac{4\pi}{qk}} .
\ea
\label{eq:J-WS-1q}
\ee
Note that these are also consistent with the planar free energy computed in appendix \ref{app:planar}.
In the limit $q\to 1$, these precisely reproduce the worldsheet instanton corrections in the ABJM Fermi-gas computed 
in \cite{HMO2}.
Also, for $q=2$, \eqref{eq:J-WS-1q}
reproduces the worldsheet instanton
coefficients of $(p,q)=(1,2)$ model found in \cite{MN2}.

\subsection{Pole cancellations}
Since we have determined some of the instanton coefficients analytically,
we can see the pole cancellations in some special limits of $(q,k)$ beyond the semi-classical approximation.
These are important non-trivial tests of our conjectures.

\paragraph{ABJM limit.} 
In the limit $q \to 1$, all of  $\alpha_n(q,k)$, $\gamma_n(q,k)$ and $\delta_n(q,k)$ are divergent.
Let us see the cancellation of these divergences for $n=1$.
We first notice that $I_1(q,k)$ behaves as
\be
\lim_{q \to 1} I_1(q,k)=\cO((q-1)^3).
\ee
Using the integral representation \eqref{eq:alpha1-int2},
one finds that the divergence of $\alpha_1(q,k)$ is given by
\be
\ba
\lim_{q\to 1^+}\alpha_1(q,k) e^{-\frac{2\mu}{q}}=-\frac{4}{\pi^2 k} \cos \( \frac{\pi k}{2} \) \biggl[ \frac{1}{2(q-1)^2}
+\( \mu+1-\frac{\pi k}{4} \cot \frac{\pi k}{2} \)\frac{1}{q-1} \\
+\mu^2+\(1-\frac{\pi k}{2}\cot \frac{\pi k}{2} \)\mu
+\frac{1}{2}+\frac{\pi^2(4-k^2)}{16}-\frac{\pi k}{2} \cot \frac{\pi k}{2} +\cO(q-1)\biggr]e^{-2\mu}.
\ea
\ee
The divergence of $\gamma_1(q,k)$ is
\be
\lim_{q\to 1}\gamma_1(q,k)=\frac{4}{\pi^2 k} \cos \( \frac{\pi k}{2} \) \left[ \frac{1}{q-1}+1+\cO(q-1) \right].
\ee
Thus we get
\be
\ba
&\lim_{q\to 1^+}\( \alpha_1(q,k) e^{-\frac{2\mu}{q}}+\mu \gamma_1(q,k) e^{-2\mu} \) \\
&\quad=-\frac{4}{\pi^2 k} \cos \( \frac{\pi k}{2} \) \biggl[ \frac{1}{2(q-1)^2}
+\( 1-\frac{\pi k}{4} \cot \frac{\pi k}{2} \)\frac{1}{q-1}\biggr]e^{-2\mu}  \\
&\quad\quad+( a_1(k) \mu^2+b_1(k) \mu+c_1(k)+\delta c_1(k))e^{-2\mu} +\cO(q-1),
\ea
\ee
where
\be
\ba
a_1(k)&=-\frac{4}{\pi^2 k}\cos \frac{\pi k}{2},\qquad\quad
b_1(k)=\frac{2}{\pi} \cos \frac{\pi k}{2} \cot \frac{\pi k}{2}, \\
c_1(k)&=\( -\frac{2}{3k}+\frac{5k}{12}+\frac{1}{\pi} \cot \frac{\pi k}{2}+\frac{k}{2} \csc^2 \frac{\pi k}{2} \) \cos \frac{\pi k}{2},
\ea
\ee
and
\be
\delta c_1(k)=-\frac{4}{\pi^2 k} \cos \( \frac{\pi k}{2} \)\(\frac{1}{2}+\frac{\pi^2(4-k^2)}{16}-\frac{\pi k}{2} \cot \frac{\pi k}{2}\)-c_1(k) .
\ee
The divergence of the $\mu$-dependent part is precisely canceled.
Furthermore, the coefficients $a_1(k)$ and $b_1(k)$ in the finite part perfectly agree with the results in the ABJM Fermi-gas \cite{HMO2}.
The divergence of the $\mu$-independent part must be canceled by $\delta_1(q,k)$.
This means that $\delta_1(q,k)$ must behave as
\be
\lim_{q\to 1^+}\delta_1(q,k)=\frac{4}{\pi^2 k} \cos \( \frac{\pi k}{2} \) \biggl[ \frac{1}{2(q-1)^2}
+\( 1-\frac{\pi k}{4} \cot \frac{\pi k}{2} \)\frac{1}{q-1}\biggr]-\delta c_1(k).
\ee
This is regarded as the constraint for $\delta_1(q,k)$.

\paragraph{$k\rightarrow 2$ limit.}
Let us consider another limit $k \to 2$.
In this limit, the coefficient $\alpha_1(q,k)$ of $e^{-\frac{2\mu}{q}}$ diverges, as shown in figure~\ref{fig:alpha1-q3}.
This divergence must be canceled by the leading worldsheet instanton correction of order $e^{-\frac{4\mu}{qk}}$.
It is easy to find
\be
\lim_{k \to 2} d_1(1,q,k)e^{-\frac{4\mu}{q k}}=\biggl[ \frac{2q}{\pi \sin \frac{\pi}{q}}\frac{1}{k-2}+\frac{2\mu+q+\pi \cot \frac{\pi}{q}}{\pi \sin \frac{\pi}{q}} 
+\cO(k-2)\biggr]e^{-\frac{2\pi}{q}}.
\ee
Using the integral expression \eqref{eq:alpha1-int} or \eqref{eq:alpha1-int2}, we also find
\be
\lim_{k \to 2} \alpha_1(q,k)e^{-\frac{2\mu}{q}}=\biggl[ -\frac{2q}{\pi \sin \frac{\pi}{q}}\frac{1}{k-2}-\frac{\cot \frac{\pi}{q}}{\sin \frac{\pi}{q}}
+\cO(k-2) \biggr]e^{-\frac{2\pi}{q}}.
\ee
The singular parts are indeed canceled as expected.
The finite part is finally given by
\be
\lim_{k \to 2} \( \alpha_1(q,k)e^{-\frac{2\mu}{q}}+ d_1(1,q,k)e^{-\frac{4\mu}{q k}} \)=\frac{2\mu+q}{\pi \sin \frac{\pi}{q}} e^{-\frac{2\pi}{q}}.
\ee
This correctly reproduces the coefficient \eqref{k2WS} of $e^{-2\mu/q}$ 
for the $(p,q,k)=(1,q,2)$ case
(see appendix \ref{sec:variousJ}).

\paragraph{$k\rightarrow 2n$ limit.}
More generally, $\alpha_1(q,k)$ has a pole at $k=2n$ (even integer).
From the integral expression \eqref{eq:alpha1-int}, we find
that $\alpha_1(q,k)$ behaves in the limit $k\to 2n$ as
\be
\lim_{k \to 2n} \alpha_1(q,k)=-\frac{1}{k-2n}\frac{2q}{\pi\sin\frac{2\pi}{q}}
\prod_{j=1}^n\frac{\sin\frac{\pi(n+j)}{nq}}{\sin\frac{\pi j}{nq}}
+\mathcal{O}(1).
\ee
This pole should be canceled by the worldsheet $n$-instanton
\be
\lim_{k \to 2n} \( \alpha_1(q,k)e^{-\frac{2\mu}{q}}+ d_n(1,q,k)e^{-\frac{4n\mu}{q k}} \)=\text{finite}.
\label{polecancel-q1}
\ee
One can see that $d_2(1,q,k)$ in \eqref{eq:J-WS-1q} indeed has the
correct pole structure at $k=4$ satisfying the condition \eqref{polecancel-q1}.
For $n\geq3$, this pole cancellation condition \eqref{polecancel-q1}
gives the constraint for a possible form of $d_n(1,q,k)$.

\section{Results on the $(p,q)$-model}
\label{sec:pq}
In this section, we give explicit results for the general $(p,q)$-model.
The basic strategy is the same as in the case of $p=1$.
We compute the WKB expansion of each membrane instanton coefficient, and then
conjecture its analytic form.
Since the WKB expansions become much more complicated than those for the $(1,q)$-model, 
it is harder to determine their analytic forms.
To fix the worldsheet instanton corrections, we use the exact results for various integral $(p,q,k)$
in appendix \ref{sec:variousJ}.

\subsection{Exact partition function for $N=2$}
We first compute the spectral zeta function at $s=2$, exactly.
It is  easy to find that $\zeta_\rho(1)$ is exactly given by
\be
\zeta_\rho(1)=\int_{-\infty}^\infty dx\, \rho(x,x)=\frac{1}{4\pi^2 k} \frac{\Gamma^2(\frac{p}{2}) \Gamma^2 (\frac{q}{2})}{\Gamma(p)\Gamma(q)}.
\label{Z-one}
\ee
Also, $\zeta_\rho(2)$ can be computed as follows,
\be
\ba
\zeta_\rho(2)&=\int dx \bra{x}\frac{1}{(2\cosh\frac{\hat{P}}{2})^p}\frac{1}{(2\cosh\frac{\hat{Q}}{2})^q}\frac{1}{(2\cosh\frac{\hat{P}}{2})^p}\frac{1}{(2\cosh\frac{\hat{Q}}{2})^q}\ket{x}\\
&=\int dx dy \bra{x}\frac{1}{(2\cosh\frac{\hat{P}}{2})^p}\ket{y}
\bra{y}\frac{1}{(2\cosh\frac{\hat{P}}{2})^p}\ket{x} \frac{1}{(2\cosh\frac{x}{2})^q}\frac{1}{(2\cosh\frac{y}{2})^q} .
\ea
\ee
By using the fact that $\bra{x}G(\hat{P})\ket{y}$ depends only on $x-y$ and shifting
the integral variable $x\to x+y$, we find
\be
\ba
\zeta_\rho(2)&=\int \frac{dx dy}{(2\pi k)^2} 
\Big|\bra{x-y}\frac{1}{(2\cosh\frac{\hat{P}}{2})^p}\ket{0} \Big|^2
\frac{1}{(2\cosh\frac{x}{2})^q}\frac{1}{(2\cosh\frac{y}{2})^q}\\
&=\int \frac{dx}{2\pi k} 
\Big|\bra{x}\frac{1}{(2\cosh\frac{\hat{P}}{2})^p}\ket{0} \Big|^2{\cal F}(q,x),
\ea
\ee
where
\be
 {\cal F}(q,x )=\int_{-\infty}^\infty \frac{dy}{2\pi k} 
\frac{1}{(2\cosh\frac{x+y}{2})^q}\frac{1}{(2\cosh\frac{y}{2})^q}
=\frac{1}{2\pi k}
\frac{\Gamma(q)^2}{\Gamma(2q)}~{}_2F_1\left(\frac{q}{2},\frac{q}{2};\hf+q;
-\sinh^2\frac{x}{2}\right).
\ee
Using the Fourier transform
\be
\bra{x}\frac{1}{(2\cosh\frac{\hat{P}}{2})^p}\ket{0}
=
\frac{1}{2\pi}
 \B \Big(\frac{p}{2}+\frac{ix}{2\pi k}, \frac{p}{2}-\frac{ix}{2\pi k}\Big),
\ee
we finally find
\be
\zeta_\rho(2)=\frac{\B(q,q)}{8\pi^3 k}
\int_{-\infty}^\infty dx\,
\B^2\Big(\frac{p}{2}+ix, \frac{p}{2}-ix\Big)
\,{}_2F_1\left(\frac{q}{2},\frac{q}{2};\hf+q;
-\sinh^2 \pi k x \right),
\label{eq:Z2-pq}
\ee
where we have rescaled the integration variable $x \to 2\pi k x$.
Note that this is an exact expression.
Recalling the relation \eqref{eq:J}, the exact partition functions for $N=1,2$ are given by
\be
Z(1,k)=\zeta_\rho(1),\qquad
Z(2,k)=-\frac{1}{2}\zeta_\rho(2)+\frac{1}{2}\zeta_\rho(1)^2.
\ee
In the case of $p=q=1$, $Z(2,k)$ correctly reduces to the exact partition function for $N=2$ in the ABJM theory, computed in \cite{Okuyama}.

For the spectral trace $\zeta_\rho(\ell)$ with 
general $\ell\in \mathbb{Z}_{>0}$,
we conjecture that 
it has a simple integral representation
\begin{align}
 \zeta_\rho(\ell)=
\int \frac{dPdQ}{2\pi\hbar}
V(Q)U(P)e^{(-1)^{\ell}i\hbar\del_Q\del_P}V(Q)U(P)
e^{(-1)^{\ell-1}i\hbar\del_Q\del_P}V(Q)U(P)\cdots 
e^{i\hbar\del_Q\del_P}V(Q)U(P).
\label{UVtrace}
\end{align}
Here, for simplicity, 
we have introduced the notation $V(Q)=(2\cosh\frac{Q}{2})^{-q}, U(P)=(2\cosh\frac{P}{2})^{-p}$,
and the derivatives in \eqref{UVtrace} act on all functions on their right.
One can show that, for $\ell=2$, \eqref{UVtrace} indeed agrees with \eqref{eq:Z2-pq}.
Although we do not have a proof of \eqref{UVtrace} for $\ell\geq3$, we have checked that this 
conjectured expression \eqref{UVtrace} correctly reproduces
the WKB expansion.

\subsection{Membrane instanton corrections}
In this subsection, we consider the membrane instanton corrections.

\paragraph{Coefficients of $e^{-n\mu}$.}
Let us first consider the coefficient of $e^{-n\mu}$ in \eqref{eq:J-M2-pq}.
The WKB expansion of $D(-n,p,q,k)$ up to $\cO(k^4)$ is given by
\be
\ba
&D(-n,p,q,k)= 1-\frac{ n^2 (n^2-1) p^2 q^2}{96 (-1+n p) (-1+n q)}(\pi k)^2 \\
&\quad +\frac{ n^3 (n^2-1)  p^3 q^3 \left(80 n-24 p-24 n^2 p-24 q-24 n^2 q+17 n p q+7 n^3 p q\right)}{92160 (-3+n p) (-1+n p) (-3+n q) (-1+n q)}(\pi k)^4+\cO(k^6).
\ea
\ee
One notices that $D(-1,p,q,k)$ does not receive the quantum corrections: $D(-1,p,q,k)=1$.
This is indeed the case. Since $\zeta_\rho(1)$ does not receive any corrections, we have $D(1,p,q,k)=1$ for any $(p,q)$.
Using the reflection symmetry \eqref{eq:D-sym}, we conclude that $D(-1,p,q,k)=1$.
Therefore $\beta_1(p,q,k)$ is exactly given by
\be
\beta_1(p,q,k)=\frac{\beta_1^{(0)}(p,q)}{k}=\frac{1}{4\pi^2 k} \frac{\Gamma^2(-\frac{p}{2})}{\Gamma(-p)}
\frac{\Gamma^2(-\frac{q}{2})}{\Gamma(-q)}.
\ee

Moreover, the WKB expansion for $n=2$ has the following remarkable structure:
\be
D(-2,p,q,k)=\sum_{n=0}^\infty  \frac{(-1)^n}{(2n)!}f_n(p)f_n(q)\( \frac{\pi k}{2} \)^{2n},
\label{D2factorize}
\ee
where a generating function of $f_n(p)$ is given by
\be
\sum_{n=0}^\infty \frac{f_n(p)}{(2n)!}z^{2n}={}_2F_1\( -\frac{p}{2},-\frac{p}{2};\frac{1}{2}-p;\sin^2 z\).
\ee
More explicitly, $f_n(p)$ is given by
\be
f_n(p)=\sum_{m=0}^n \frac{(-1)^n}{2^{2m-1}m!} \frac{(-\frac{p}{2})_m^2}{(\frac{1}{2}-p)_m} \sum_{j=1}^m (-1)^j \binom{2m}{m-j}(2j)^{2n}.
\ee
In the previous subsection, we have already computed $\zeta_\rho(2)$.
From \eqref{eq:Z2-pq}, one finds that
\be
D(2,p,q,k)=\frac{\Gamma(2p)}{2\pi \Gamma^2(p)}\int_{-\infty}^\infty dx\,
\B^2\Big(\frac{p}{2}+ix, \frac{p}{2}-ix\Big)
\,{}_2F_1\left(\frac{q}{2},\frac{q}{2};\hf+q;
-\sinh^2 \pi k x \right).
\ee
Using the symmetry \eqref{eq:D-sym}, one immediately obtains
\be
\ba
&D(-2,p,q,k)=D(2,-p,-q,k) \\
&=\frac{\Gamma(-2p)}{2\pi \Gamma^2(-p)}\int_{-\infty}^\infty dx\,
\B^2\Big(-\frac{p}{2}+ix, -\frac{p}{2}-ix\Big)
\,{}_2F_1\left(-\frac{q}{2},-\frac{q}{2};\hf-q;
-\sinh^2 \pi k x \right).
\ea
\ee
One can understand the factorized structure \eqref{D2factorize} of $D(-2,p,q,k)$ 
from the expression of $\zeta_\rho(2)$ in \eqref{UVtrace}.
We emphasize that this expression is valid for any $(p,q)$.
In particular, for $p=1,2,3,4$, we find the following analytic expressions
\be
\ba
D(-2,1,q,k)&={}_2F_1\left(-\frac{q}{2},-\frac{q}{2},\hf-q;\sin^2\frac{\pi k}{2}\right),\\
D(-2,2,q,k)&=\frac{2}{3}+\frac{1}{3}\,{}_2F_1\left(-\frac{q}{2},-\frac{q}{2},\hf-q;\sin^2\pi k\right),\\
D(-2,3,q,k)&=\frac{9}{10}
\,{}_2F_1\left(-\frac{q}{2},-\frac{q}{2},\hf-q;\sin^2\frac{\pi k}{2}\right)
+\frac{1}{10}\,{}_2F_1\left(-\frac{q}{2},-\frac{q}{2},\hf-q;\sin^2\frac{3\pi k}{2}\right),\\
D(-2,4,q,k)&=\frac{18}{35}+\frac{16}{35}\,{}_2F_1\left(-\frac{q}{2},-\frac{q}{2},\hf-q;\sin^2\pi k\right)
+\frac{1}{35}\,{}_2F_1\left(-\frac{q}{2},-\frac{q}{2},\hf-q;\sin^22\pi k\right).
\label{eq:D(-2)}
\ea
\ee
Similarly, if either $D(n,p,q,k)$ or $D(-n,p,q,k)$ ($n \in \mathbb{Z}_{>0}$) is known,
one can know the other by the symmetry \eqref{eq:D-sym} and the analytic continuation $(p,q) \to (-p,-q)$.
By matching the WKB data, we find the analytic form of $D(-n,p,q,k)$ for some other cases
\be
\ba
 D(-3,2,q,k)=&\frac{3}{5} \, _2F_1\left(-q,-\frac{q}{2};\frac{1}{2}-\frac{3 q}{2};\sin ^2 \pi k\right)+\frac{2}{5},\\
 D(-3,4,q,k)=&\frac{8}{77} \, _2F_1\left(-\frac{3 q}{2},-q;\frac{1}{2}-\frac{3 q}{2};\sin ^2\pi k\right)+\frac{48}{77} \,
   _2F_1\left(-q,-\frac{q}{2};\frac{1}{2}-\frac{3 q}{2};\sin ^2\pi k\right) \\
   &+\frac{3}{77} \, _2F_1\left(-q,-\frac{q}{2};\frac{1}{2}-\frac{3 q}{2};\sin
   ^2 2  \pi k\right)+\frac{18}{77},\\
   D(-4,2,q,k)=&
   \frac{16}{35} \, _2F_1\left(-3 q,-q;\frac{1}{2}-2 q;\sin ^2\frac{\pi k}{2}\right)+\frac{2}{35} \, _2F_1\left(-2 q,-q;\frac{1}{2}-2 q;\sin
   ^2\pi k\right)\\
   &+\frac{9}{35} \, _2F_1\left(-q,-q;\frac{1}{2}-2 q;\sin ^2\pi k\right)+\frac{8}{35}.
\label{eq:D(-3)}
\ea
\ee
Note that when $q$ is an integer, the hypergeometric series in \eqref{eq:D(-2)}
and \eqref{eq:D(-3)} are truncated to a finite sum, and they are reduced to
some combinations of trignometric functions.

\paragraph{Coefficients of $e^{-2n\mu/p}$.}
The WKB expansion of the coefficient of $e^{-2n\mu/p}$ up to $\cO(k^4)$ is 
\be
\ba
&D\(-\frac{2n}{p},p,q,k\)=1-\frac{n^2 q^2 (2 n-p) (2 n+p)}{24 p (-1+2 n) (-p+2 n q)}(\pi k)^2 \\
&\qquad\quad+\frac{n^3 q^3 (2 n-p) (2 n+p)}
{5760p^2 (-3+2 n) (-1+2 n)  (-3 p+2 n q) (-p+2 n q)}\bigl(80 n p-48 n^2 p\\
&\qquad\qquad\qquad\qquad-12 p^3-48 n^2 q+28 n^3 q-12 p^2 q+17 n p^2 q\bigr)(\pi k)^4+\cO(k^6).
\ea
\label{eq:M2-fractional-WKB}
\ee
Here we focus on the $n=1$ case, i.e., the coefficient of $e^{-2\mu/p}$.
For the case of $q/p=m\in\mathbb{Z}_{>0}$,
we find that the following expression correctly reproduces the WKB expansion
up to $\mathcal{O}(k^{34})$
\begin{align}
 D\(-\frac{2}{p},p,mp,k\)=
\frac{m!^2}{(2m)!}\sum_{\sum_j j\lambda_j=m}\prod_{j=1}^m\frac{1}{\lambda_j!}
\left[\frac{(-1)^{j-1}}{j}\frac{mp\sin\pi jk}{\sin\frac{\pi jkp}{2}}\right]^{\lambda_j}.
\end{align}
One can easily show that this can be rewritten as a contour integral
\begin{align}
  D\(-\frac{2}{p},p,mp,k\)=&
\frac{\Gamma(m+1)^2}{\Gamma(2m+1)}\oint_{z=0}
\frac{dz}{2\pi iz^{m+1}}\exp\left[mp\sum_{j=1}^\infty \frac{(-1)^{j-1}}{j}\frac{\sin\pi jk}{\sin\frac{\pi jkp}{2}}\right]\nn
=&
\frac{\Gamma(m+1)^2}{\Gamma(2m+1)}\oint_{z=0}
\frac{dz}{2\pi iz^{m+1}}\left[\prod_{n=0}^\infty \frac{1+z\mathfrak{q}^{n-\frac{1}{p}+\hf}}{1+z\mathfrak{q}^{n+\frac{1}{p}+\hf}}\right]^{mp},
\label{r1zoint}
\end{align}
where $\mathfrak{q}=e^{i\pi kp}$.
In particular, when $(p,q)=(1/m,1)$, this integral can be evaluated explicitly
by expanding the integrand using the $q$-binomial formula.
By picking up the coefficient of $z^m$, we find
\begin{align}
D\(-2m,\frac{1}{m},1,k\)=
\frac{m!^2}{(2m)!}
\frac{\prod_{j=1}^{2m}\sin\frac{\pi kj}{2m}}{\prod_{j=1}^{m}\sin^2\frac{\pi kj}{2m}}.
\label{qfactorial}
\end{align}
By noticing that \eqref{qfactorial} can be written as a combination
of $q$-factorials,
we have arrived at the conjecture in the previous section that
the coefficient of $e^{-2\mu/p}$ is given by
the $q$-gamma function \eqref{r1qGamma}  for the general $(p,1)$ case.
Also, for the $(p,q)=(2/m,2)$ case, 
the integral \eqref{r1zoint} can be evaluated 
exactly thanks to the formula (3.9.1)
in \cite{koekoek}
\begin{align}
  D\(-m,\frac{2}{m},2,k\)
&=
\frac{m!^2}{(2m)!}{}_2\phi_1(\mathfrak{q}^{-m},\mathfrak{q}^{-m},\mathfrak{q};
\mathfrak{q},\mathfrak{q}^{m+1})\nn
&=
\frac{m!^2}{(2m)!}\sum_{l=0}^m \frac{\prod_{j=1}^m\sin^2\frac{\pi kj}{m}}{
\prod_{j=1}^l\sin^2\frac{\pi kj}{m}\prod_{j=1}^{m-l}\sin^2\frac{\pi kj}{m}},
\label{qhyp}
\end{align}
where ${}_2\phi_1(\mathfrak{q}^{a},\mathfrak{q}^{a},\mathfrak{q}^c;
\mathfrak{q},z)$ denotes the basic $q$-hypergeometric series.
This suggests that the 
the coefficient of $e^{-2\mu/p}$ for the general $(p,2)$ case
is given by a certain analytic continuation
of \eqref{qhyp} to a non-integer $m$.
However, compared to  the 
$q$-gamma function appearing in the $(p,1)$ case,
the precise definition of the $q$-hypergeometric series with
$|\mathfrak{q}|=1$ is much more subtle (see \cite{nishizawa} for some proposal).
We leave it as an interesting future problem.
For $q>2$, the integral \eqref{r1zoint} is hard to evaluate explicitly.

When $m=q/p$ is not integer, the contour integral representation
\eqref{r1zoint} is no longer correct due to the branch cut of $1/z^{q/p+1}$.
Instead, we conjecture that the  coefficient of $e^{-2\mu/p}$
for the general $(p,q)$ case is given by the following integral along the real segment $0<z<1$
\be
\ba
D\(-\frac{2}{p},p,q,k\)=1+\frac{2\Gamma(-2q/p)}{\Gamma^2(-q/p)} 
\int_0^1 \frac{dz}{z^{1+q/p}} \biggl( e^{q F(p,k,z)} -(1+z)^{2q/p} \biggr),
\ea
\label{eq:M2-fractional}
\ee
with
\be
F(p,k,z)=2\sum_{n=0}^\infty \frac{(-1)^{n-1}}{(2n+1)!}B_{2n+1}\(\frac{1}{2}+\frac{1}{p}\)\Li_{1-2n}(-z) (\pi k p)^{2n}.
\label{eq:F-sum}
\ee
Here $\Li_n(z)$ denotes the polylogarithm.
We do not have a proof of this conjecture, but we have confirmed that it correctly reproduces the WKB expansion up to $\cO(k^{34})$.
For $p\geq 2$, one can rewrite $F(p,k,z)$ as the following integral form by using \eqref{eq:Bernoulli-int},
\be
F(p,k,z)=\int_0^\infty dt \frac{\sin \frac{2\pi}{p}}{\cosh 2\pi t+\cos \frac{2\pi}{p}} \log ( 1+2z \cosh \pi k p t+z^2),\qquad p\geq 2.
\label{eq:F-int}
\ee
For $p<2$, one has to shift the argument of the Bernoulli polynomial in \eqref{eq:F-sum} to use the integral representation.

\subsection{Worldsheet instanton corrections}
As in the same way in the previous section, we can compute the worldsheet instanton
 corrections for various integral $(p,q ,k)$
(see appendix~\ref{sec:variousJ}).
From these data, we conjecture that the leading worldsheet instanton 
correction in \eqref{eq:J-WS}
is given by
\be
d_1(p,q,k)=\frac{pq}{\sin \frac{2\pi}{pk} \sin \frac{2\pi}{q k}}.
\label{WS1inst-pq}
\ee
To guess the higher order corrections is not easy.
For the $(1,q)$-model, we conjectured the 2-instanton correction in \eqref{eq:J-WS-1q}.
We also conjecture the 2-instanton correction
for the $(2,q)$-model as
\be
d_2(2,q,k)=-\frac{2}{\sin^2\frac{2\pi}{qk}}
-\frac{q^2}{\sin^2\frac{\pi}{k}}+\frac{q(2+\cos\frac{4\pi}{qk})}{\sin\frac{2\pi}{k}\sin\frac{4\pi}{qk}}.
\label{WS2inst-2q}
\ee
One can check that in the limit $q\to2$ this reproduces the result of $(p,q)=(2,2)$
model in \cite{MN2}.

\subsection{Pole cancellations}

The worldsheet 1-instanton coefficient \eqref{WS1inst-pq} for the general
$(p,q)$ case has a pole at $k=2/q$
\begin{align}
 \lim_{k\to \frac{2}{q}}d_1(p,q,k)e^{-\frac{4\mu}{pqk}}
 =\frac{1}{\pi\sin\frac{\pi q}{p}}\left[\frac{2p}{k-\frac{2}{q}}+\pi q^2\cot\frac{\pi q}{p}+q(2\mu+p)\right]e^{-\frac{2\mu}{p}}.
 \label{ws1inst-pole}
\end{align}
This pole should be canceled by the membrane 1-instanton coefficient
 $\al_1(p,q,k)$ of
$e^{-2\mu/p}$
given by \eqref{eq:M2-fractional}.
We have checked numerically that the membrane 1-instanton coefficient \eqref{eq:M2-fractional} has the correct pole and residue to cancel the pole of worldsheet instanton \eqref{ws1inst-pole}.
Similarly, the pole of the membrane 1-instanton at $k=4/q$ should be canceled
by the worldsheet 2-instanton
\begin{align}
 \lim_{k\to \frac{4}{q}}\left[d_2(p,q,k)e^{-\frac{8\mu}{pqk}}+\al_1(p,q,k)e^{-\frac{2\mu}{p}}\right]=\text{finite}.
\end{align}
This gives the constraint for a possible form of $d_2(p,q,k)$.
However, it is difficult to numerically calculate 
the integral \eqref{eq:M2-fractional} in the regime $k\geq2/q$, and 
hence we are unable to determine the residue
 of $\al_1(p,q,k)$ at $k=4/q$ so far. Also, it is not clear whether
 the residue of $\al_1(p,q,k)$ at $k=4/q$ for the $q\geq3$ case
 is simply given by trigonometric functions.
 It would be interesting to find the exact form of worldsheet
 2-instanton coefficient $d_2(p,q,k)$
 for the general $(p,q)$ case. We leave it as a future problem.

\section{More results in special cases}
\label{sec:special}
In this section,
we discuss results for some specific values of $(p,q)$.
In some special cases, there is a direct connection to the topological strings on certain Calabi-Yau three-fold.

\subsection{Relation to the topological strings}
\subsubsection{The $(2,2)$-model and the local  $D_5$ del Pezzo}
In \cite{MN3}, it was observed that the worldsheet instanton correction in the $(2,2)$-model can be
reproduced by the topological string on the local $D_5$ del Pezzo surface.
Here we show that the Fermi surface \eqref{eq:Fermi-surface} with $p=q=2$ is indeed equivalent
to the mirror curve for the $D_5$ del Pezzo.
Let us rewrite \eqref{eq:Fermi-surface} as
\be
(e^{P/2}+e^{-P/2})^p(e^{Q/2}+e^{-Q/2})^q=e^E.
\label{eq:Fermi-surface-2}
\ee
For $p=q=2$, this reduces to
\be
e^{P+Q}+e^P+e^{P-Q}+2e^{Q}+4+2e^{-Q}+e^{-P+Q}+e^{-P}+e^{-P-Q}=e^E.
\ee
Looking at fig.~1 in \cite{HKP}, one finds that this Fermi surface is identical to 
the mirror curve for the $D_5$ del Pezzo.\footnote{%
The local $D_5$ del Pezzo surface corresponds to the polyhedron $15$ in fig.~1 in \cite{HKP}.
}
Following the formulation in \cite{ACDKV, GHM1}, the (quantized) mirror curve is enough to compute the free energy
on the corresponding geometry in the topological string.
However, one should be careful about the prescription of the quantization of the mirror curve. 
As in \cite{GHM1}, a natural way to qunatize the mirror curve is Weyl's prescription:
\be
e^{r P+ s Q} \to e^{r \hat{P}+s \hat{Q}}.
\ee 
On the other hand, the quantization of the Fermi surface \eqref{eq:Fermi-surface-2} leads to
\be
(e^{\hat{P}/2}+e^{-\hat{P}/2})^p(e^{\hat{Q}/2}+e^{-\hat{Q}/2})^q\ket{\psi}=e^E \ket{\psi},
\label{eq:quantum-Fermi-surface}
\ee
where $\ket{\psi}$ is an eigenstate in the quantum spectral problem.
This quantization induces an additional factor:
\be
e^{r \hat{P}}e^{s \hat{Q}}=e^{- rs\pi i k}e^{r \hat{P}+s \hat{Q}}.
\ee
Such $k$-dependent factors should be taken into account appropriately when one computes the membrane instanton correction
from the topological string free energy.%
\footnote{Note that the worldsheet instanton correction is determined by the classical mirror curve.
We do not need the quantization of the curve in the computation.}
In particular, one should carefully indentify the moduli in the topological string and the parameters in the Fermi-gas system.
See \cite{KM} in more detail in the ABJM case.

\subsubsection{The $(1,-1)$-model and the resolved conifold}
For $(p,q)=(1,-1)$,\footnote{%
This case cannot be defined in the original setup.
We define this case as a naive analytic continuation in the ideal Fermi gas system.
}  the function $D(s,1,-1,k)$ becomes remarkably simple:
\begin{\eq}
D(s,1,-1,k) = \frac{\pi k s}{4\sin{\frac{\pi ks}{4}}} .
\end{\eq}
Then we can compute the WKB grand potential by
\begin{\eq}
\mathcal{J}_{\rm WKB}(\mu ,k)
= -\frac{\pi}{4}\int_{c-i\infty}^{c+i\infty} \frac{ds}{2\pi i}
\frac{e^{s\mu}}{s\sin{\frac{\pi ks}{4}} \sin^2{\frac{\pi s}{2}}}  ,
\end{\eq}
where we have used $\Gamma (z)\Gamma(-z)=-\pi/(z\sin{(\pi z)})$.
By taking the integral contour $\mathcal{C}_+$ in fig.~\ref{fig:contour}
and picking the poles at $s=4n/k$ and $s=2n$  with $n\in\mathbb{Z}$,
we find
\begin{\eq}
\mathcal{J}_{\rm WKB}(\mu ,k)
=\frac{1}{4}\sum_{n=1}^\infty 
\frac{(-1)^n e^{\frac{4n\mu}{k}}}{n\sin^2{\frac{2\pi n}{k}}}
+\frac{1}{4\pi}\sum_{n=1}^\infty \frac{1}{n^2 \sin{\frac{\pi kn}{2}}}
\Biggl[ 2n\mu -1 -\frac{\pi k}{2} \cot{\frac{\pi kn}{2}} \Biggr] e^{2n\mu} .
\label{eq:J_p1m1}
\end{\eq}
The first term is similar to the WS instanton effects
while the second terms is similar to the membrane instantons.
Although each term has poles for rational values of $k$,
these poles are actually canceled and the result is finite. 

We can understand 
this expression from the refined topological string on the resolved conifiold as follows.
First let us note that
the classical Fermi surface for $(p,q)=(1,-1)$ is determined by
\begin{\eq}
\frac{e^{\frac{P}{2}} +e^{-\frac{P}{2}}}{e^{\frac{Q}{2}} +e^{-\frac{Q}{2}} } =e^E  .
\end{\eq}
\begin{\eq}
1 +e^{Q '} +e^{P'} -e^{-2E}e^{Q'-P'} =0 ,
\end{\eq}
where $(Q',P')=(Q,\frac{Q+P}{2})$.
This equation is the same as the mirror curve of the resolved conifold (see \cite{HKRS} for instance) and
hence we expect that the $(p,q)$-model for $(p,q)=(1,-1)$ is described
by the topological string on the resolved conifold.

Let us explicitly test our expectation (see also \cite{Hatsuda}).
The free energy of the refined topological string on the resolved conifold is given by \cite{HKRS}
\begin{\eq}
F(\epsilon_1 ,\epsilon_2 ;Q )
=-\sum_{n=1}^\infty \frac{Q^n}{n(q^{n/2}-q^{-n/2})(t^{n/2}-t^{-n/2})} ,
\end{\eq}
where
\begin{\eq}
q=e^{\epsilon_1},\quad t=e^{-\epsilon_2}.
\end{\eq}
Then the Nekrasov-Shatashvili limit \cite{NS} becomes
\begin{\eq}
F_{\rm NS}(\epsilon_1 ;Q)
= \lim_{\epsilon_2 \rightarrow 0} \epsilon_2 F (\epsilon_1 ,\epsilon_2 ;Q)
=\sum_{n=1}^\infty \frac{1}{n^2} \frac{Q^n}{q^{n/2}-q^{-n/2}} .
\end{\eq}
If we identify the parameters as
\begin{\eq}
\epsilon_1 =\pi ik ,\quad Q=e^{\frac{T}{\lambda}},\quad T= \frac{4\mu}{k},\quad \lambda =\frac{2}{k} ,
\end{\eq}
then we find
\begin{\eq}
F_{\rm NS}(\lambda ,T )
=\frac{1}{2i} \sum_{n=1}^\infty \frac{e^{\frac{nT}{\lambda}}}{n^2 \sin{\frac{\pi n}{\lambda}}} .
\label{\eq:NS_p1m1}
\end{\eq}
Also, in the standard topological string limit $\epsilon_1 = -\epsilon_2$ with the identifications
\begin{\eq}
\epsilon_1 =\frac{4\pi i}{k} ,\quad Q=e^{\frac{4\mu}{k}+\pi i} ,
\end{\eq}
the free energy becomes
\begin{\eq}
F_{\rm top} (k,\mu)
= \frac{1}{4}\sum_{n=1}^\infty \frac{(-1)^n e^{\frac{4n\mu}{k}}}{n\sin^2{\frac{2\pi n}{k}}} .
\label{eq:top_p1m1}
\end{\eq}
By comparing the grand potential \eqref{eq:J_p1m1}
with \eqref{\eq:NS_p1m1} and \eqref{eq:top_p1m1},
we easily see 
\begin{\eq}
\mathcal{J}_{\rm WKB}(\mu ,k)
=F_{\rm top} (k,\mu) 
+\frac{1}{2\pi i}\frac{\del}{\del\lambda}\Biggl[ \lambda F_{\rm NS}(\lambda ,T ) \Biggr] .
\end{\eq}
This structure is the same as the relation between
the ABJ(M) theory and refined topological string on local $\mathbb{P}^1 \times\mathbb{P}^1$ \cite{HMMO,HoO,MaMo}.

\subsubsection{A comment on general case}
For general values of $(p,q)$, we do not find the correspondence to the topological strings on the known Calabi-Yau
geometries.
As in section~\ref{sec:Fermi}, the WKB expansions of the membrane instanton corrections can be computed for any $(p,q)$
even though we do not know the topological string counterpart.
Let us give a comment how to compute the worldsheet instanton corrections systematically for generic $(p,q)$.
As seen above, the Fermi surface is closely related to the mirror curve of the corresponding topological string.
This suggests us to regard the Fermi surface \eqref{eq:Fermi-surface-2} for general $(p,q)$ as
a ``mirror curve'' of an unconvetional geometry.
Using the formulation in \cite{BKMP}, one can, in principle, compute the genus $g$ free energy for this ``mirror curve.''
It is natural to expect that this free energy just gives the worldsheet instanton corrections in the Fermi-gas system.
The important point is that the formulation in \cite{BKMP} can be applied for any spectral curve even if its geometrical meaning
is unclear.
In practice, however, it is not easy to compute the higher genus correction in this way.
It would be interesting to test this expectation explicitly.

\subsection{Exact partition function for the $(2,2)$-model at $k=1$}
The $(2,2)$-model was studied in \cite{MN3} in detail.
Here we point out that the grand potential at $k=1$ is exactly related to the topological string free energy on local $\mathbb{P}^1 \times \mathbb{P}^1$.%
\footnote{As already seen, the $(2,2)$-model is related to the topological string on the local $D_5$ del Pezzo surface.
The relation here is probably accidental.
}
The modified grand potential for the $(2,2)$-model at $k=1$ is given by (see \cite{MN3} for detail)
\be
J_{2,2}(\mu,1)=\frac{\mu^3}{6\pi^2}-\frac{2\zeta(3)}{\pi^2}+\frac{4\mu^2+4\mu+4}{\pi^2}e^{-\mu}
+\biggl[ -\frac{26\mu^2+\mu+9/2}{\pi^2}+2 \biggr]e^{-2\mu}+\cdots.
\label{eq:J22}
\ee
One notices that this large $\mu$ expansion is very similar to the one in the ABJM Fermi-gas at $k=2$ \cite{HMO2}.
In \cite{CGM}, the ABJ(M) grand potential at $k=1,2$ is exactly written in terms of the topological string free energy.
Recalling this fact, one easily finds that the modified grand potential \eqref{eq:J22} is written as
\be
J_{2,2}(\mu,1)=\frac{1}{\pi^2}\( F_0(t)-t F_0'(t)+\frac{t^2}{2}F_0''(t)\)+F_1(t)+F_1^\text{NS}(t).
\label{eq:J22-exact}
\ee
Several definitions are in order.
The functions $F_0(t)$ and $F_1(t)$ are the standard genus zero and genus one free energies on local $\mathbb{P}^1 \times \mathbb{P}^1$, respectively.%
\footnote{%
Since there are two K\"ahler moduli $t_1$ and $t_2$ in local $\mathbb{P}^1 \times \mathbb{P}^1$,
the free energy is in general a function of these two parameters $(t_1,t_2)$.
Here we denote the free energy in the diagonal slice by $F_g(t,t)=F_g(t)$.
}
These are computed in a standard way of the special geometry.
The function $F_1^\text{NS}(t)$ is the first correction to the refined topological string free energy in the Nekrasov-Shatashvili limit.
The K\"ahler modulus $t$ is related to the complex modulus $z$ by the mirror map
\begin{align}
t=-\log z+4z\, {}_4F_3\(1,1,\frac{3}{2},\frac{3}{2};2,2,2;-16z\). 
\label{eq:mirrormap}
\end{align}
In the present case, the complex modulus $z$ is related to the chemical potential or fugacity by
\begin{align}
z=e^{-\mu}=\frac{1}{\kappa}.
\end{align}
As in \cite{CGM}, the genus zero free energy is written in the closed form
\be
F_0''(t)=\pi \frac{\mathbf{K}(1+16z)}{\mathbf{K}(-16z)}+\pi i.
\ee
At the large radius point ($t \to \infty$), this leads to
\begin{align}
F_0(t)=\frac{t^3}{6}-2\zeta(3)+4e^{-t}-\frac{9}{2}e^{-2t}+\frac{328}{27}e^{-3t}-\frac{777}{16}e^{-4t}+\cO(e^{-5t}),
\end{align}
where we have fixed integration constants properly by following \cite{CGM}.
Eliminating $t$ by \eqref{eq:mirrormap}, one easily finds
\begin{align}
&F_0(t)-t F_0'(t)+\frac{t^2}{2}F_0''(t) =-\frac{1}{6}\log^3 z-2\zeta(3)+4(\log^2 z-\log z+1)z \notag \\
&\qquad +\(-26\log^2 z+\log z-\frac{9}{2}\)z^2+\frac{8}{27}(828\log^2 z+228 \log z+77)z^3+\cO(z^4).
\end{align} 
The free energies $F_1(t)$ and $F_1^\text{NS}(t)$ are also exactly given by
\begin{align}
F_1(t)&=-\frac{1}{12}\log [ 64z(1+16z) ]-\frac{1}{2} \log \( \frac{\mathbf{K}(-16z)}{\pi} \)\\
&=-\frac{\log z}{12}+\frac{2}{3}z-\frac{10}{3}z^2+\frac{224}{9}z^3+\cO(z^4), \notag \\
F_1^\text{NS}(t)&=\frac{1}{12}\log z-\frac{1}{24}\log(1+16z) \\
&=\frac{\log z}{12}-\frac{2}{3}z+\frac{16}{3}z^2-\frac{512}{9}z^3+\cO(z^4). \notag
\end{align}
Plugging these results into \eqref{eq:J22-exact}, one can check that the large $\mu$ expansion \eqref{eq:J22}
is correctly reproduced.

\paragraph{Exact grand partition function.}
Once the modified grand potential is known, one can compute the grand partition function.
The grand partition function is constructed from the modified grand potential by
\begin{align}
\Xi(\mu,k)=\sum_{n \in \mathbb{Z}} e^{J(\mu+2\pi i n,k)}.
\end{align}
Plugging the result \eqref{eq:J22-exact} into the summand in this equation, one finds
\begin{align}
e^{J_{2,2}(\mu+2\pi i n,1)}=e^{J_{2,2}(\mu,1)}\exp \left[ \pi i n^2 \tau+2\pi i n\( \xi-\frac{2}{3} \) \right],
\end{align}
where
\begin{align}
\tau=\frac{2i}{\pi} F_0''(t),\qquad
\xi=\frac{1}{\pi^2}(tF_0''(t)-F_0'(t)).
\end{align}
We have used the identity $\exp \bigl[ -\frac{4\pi i n^3}{3} \bigr]=\exp \bigl[- \frac{4\pi i n}{3} \bigr]$ with $n \in \mathbb{Z}$.
Therefore the exact grand partition function is expressed in terms of the Jacobi theta function
\begin{align}
\Xi_{2,2}(\mu,1)=e^{J_{2,2}(\mu,1)} \vartheta_3 \( \xi-\frac{2}{3},\tau \).
\label{eq:Xi22}
\end{align}
This expression is useful in $\mu \to \infty$. Now we want to analytically continue it
to the regime $\mu \to -\infty$ (or $\kappa \to 0$).
To do so, we write $\Xi_{2,2}$ in terms of periods around the orbifold point \cite{DMP1, CGM}:
\be
\ba
\lambda&= \frac{\tilde{\kappa}}{8\pi}\, {}_3F_2 \( \frac{1}{2},\frac{1}{2},\frac{1}{2};1,\frac{3}{2};-\frac{\tilde{\kappa}^2}{16}\), \\
\pd_\lambda \cF_0(\lambda)&= \frac{\tilde{\kappa}}{4} 
G^{2,3}_{3,3}\biggl( \begin{array}{c} \frac{1}{2},\frac{1}{2},\frac{1}{2}\\ 0,0,-\frac{1}{2} \end{array}\bigg| -\frac{\tilde{\kappa}^2}{16} \biggr)
+\frac{\pi^2 i \tilde{\kappa}}{2} \,{}_3F_2 \( \frac{1}{2},\frac{1}{2},\frac{1}{2};1,\frac{3}{2};-\frac{\tilde{\kappa}^2}{16}\),
\ea
\ee
where $\kappa=\tilde{\kappa}^2$ and $G^{m,n}_{p,q}$ is the Meijer G-function. 
Along the computation in \cite{CGM}, one finds that the grand partition function is finally given by
\be
\Xi_{2,2}(\mu,1)=\exp \left[ -\frac{4}{\pi^2} \( \cF_0 -\lambda \pd_\lambda \cF_0+\frac{\lambda^2}{2} \pd_\lambda^2 \cF_0 \)
+\cF_1+F_1^\text{NS} \right] \vartheta_2 (\bar{\xi}, \bar{\tau} ),
\ee
where
\be
\bar{\tau}=-\frac{1}{\tau}=\frac{i}{8\pi^3} \pd_\lambda^2 \cF_0 ,\quad
\bar{\xi}=\frac{\xi-\frac{1}{6}}{\tau}=\frac{i}{2\pi^3}( \lambda \pd_\lambda^2 \cF_0-\pd_\lambda \cF_0),
\ee
and
\be
\cF_1=-\log \eta(2\bar{\tau})-\frac{1}{2} \log 2.
\ee
Now, we can expand the grand partition function around $\kappa=\tilde{\kappa}^2=0$.
Using the expasions
\be
\ba
&-\frac{4}{\pi^2} \( \cF_0 -\lambda \pd_\lambda \cF_0+\frac{\lambda^2}{2} \pd_\lambda^2 \cF_0 \)
=\frac{\kappa}{8\pi^2}-\frac{\kappa^2}{384\pi^2}+\frac{199\kappa^3}{2211840\pi^2}+\cO(\kappa^4), \\
&\cF_1+F_1^\text{NS}=-\frac{\log \kappa}{8}+\frac{\kappa^2}{32768}-\frac{\kappa^3}{524288}+\cO(\kappa^4), \\
&\log \vartheta_2(\bar{\xi},\bar{\tau} )=\frac{1}{8} \log \kappa 
+\frac{\kappa }{8 \pi ^2}-\frac{1536-256 \pi ^2+9 \pi ^4 }{294912 \pi ^4}\kappa ^2 \\
&\hspace{2.5cm}+\frac{368640+614400 \pi ^2-128288 \pi ^4+2025 \pi ^6 }{1061683200 \pi ^6}\kappa ^3
+\cO(\kappa^4),
\ea
\ee
we finally get
\be
\Xi_{2,2}(\mu,1)=1+\frac{\kappa }{4 \pi ^2}+\frac{15-\pi ^2 }{576 \pi ^4}\kappa ^2
+\frac{855+75 \pi ^2-16 \pi ^4 }{518400 \pi ^6}\kappa ^3
+\cO(\kappa^4).
\ee
This precisely reproduces the exact values of the partition function in \cite{MN3}.

\section{Conclusions}
\label{sec:con}
In this paper we have studied the partition function of the $(p,q)$-model on $S^3$
and investigated its large $N$ instanton effects by using the Fermi-gas approach. 
Based on the systematic semi-classical WKB analysis, 
we have found the analytic results on the membrane instanton corrections.
The membrane instanton coefficient of the type $e^{-n\mu}$ is related to the
spectral zeta function $\zeta_\rho(n)$ by the reflection symmetry \eqref{eq:D-sym}.
From the explicit forms of $\zeta_\rho(1)$ in \eqref{Z-one}
and $\zeta_\rho(2)$ in \eqref{eq:Z2-pq},
we know the exact expressions of the coefficient of $e^{-n\mu}$ for $n=1,2$.
As shown in \eqref{eq:D(-2)} and \eqref{eq:D(-3)},
when $p$ is an integer with  generic $q$, the coefficients of $e^{-n\mu}$ reduce
to some combinations of the hypergeometric functions.
The membrane instanton of the type $e^{-2n\mu/p}$ (or $e^{-2n\mu/q}$) is more involved.
We found an integral representation \eqref{eq:M2-fractional}
of the coefficient of 1-instanton $e^{-2\mu/p}$
for generic $(p,q,k)$. Very surprisingly,
for the special case of $(p,q)=(1,q)$, the coefficient of
$e^{-2\mu/q}$ is given by the $q$-gamma functions \eqref{r1qGamma}.
We emphasize that
this is quite different from the Gopakumar-Vafa type formula \cite{Gopakumar:1998jq} in topological string, 
where only trigonometric functions of $\hbar$ or $1/\hbar$ appear.
It is very interesting to understand the physical meaning of this finding better.
From the observation of the 
special case \eqref{qhyp},
we speculate that for the general $(p,q)$ case, the coefficient of $e^{-2\mu/p}$ in \eqref{eq:M2-fractional}
is related to $q$-hypergeometric functions.

We have also found some exact results on worldsheet instanton corrections, 
which appear as the quantum mechanical non-perturbative corrections in the Fermi gas,
from the exact computation of the partition functions at finite $N$.
We have found the worldsheet 1-instanton for the general $(p,q)$ case in \eqref{WS1inst-pq},
and the worldsheet 2-instanton for the $(1,q)$ and $(2,q)$ cases
in \eqref{eq:J-WS-1q} 
and \eqref{WS2inst-2q}, respectively.
It would be interesting to understand more general structure of the worldsheet 
instanton corrections for the general $(p,q)$ case.

We have seen that the apparent poles at the various integral (or rational) values of $(p,q,k)$
are actually canceled out between 
the worldsheet instantons and membrane instantons, as required.
In particular, for the $(p,q)=(1,1)$ case,
after the pole cancellation  the remaining 
finite part reproduces the known results of the ABJM theory
in the highly non-trivial way.
It is interesting that the quadratic polynomial of $\mu$
in front of $e^{-2\mu}$ for the membrane instanton
of ABJM theory correctly appears from 
the $(p,q)$-model after the pole cancellation.
However, this is very mysterious from the viewpoint of bound states.
In the case of the ABJM theory, one can remove
the effects of the bound states by introducing
the effective chemical potential $\mu_\text{eff}$,
which is determined by the
coefficients of $\mu^2$ in the membrane instantons.
However, before the pole cancellation there is no $\mu^2$
term in the membrane instantons.
Therefore it seems that
there is no natural way to introduce
$\mu_\text{eff}$ in the $(p,q)$-model for generic $(p,q)$.
It would be very interesting to study the structure of the bound states in the $(p,q)$-model.

\acknowledgments{
We would like to thank Marcos Mari\~no, Sanefumi Moriyama, Tomoki Nosaka,
 Satoru Odake for useful discussions.
We are also grateful to Satoru Odake for
allowing us to use computers in the theory group, Shinshu University.
The work of KO was supported in part by JSPS Grant-in-Aid for 
Young Scientists (B)  23740178.
}

\appendix
\section{Computing the Wigner transform}\label{sec:Wigner}
In this appendix, we derive \eqref{eq:O_W}.
The computation is almost the same as the one in \cite{Hatsuda}.
By definition, the Wigner transform of $\hat{\cO}$ is given by
\be
\ba
\cO_\text{W}(Q,P)=\int_{-\infty}^\infty 
\frac{dQ' }{\hbar} 
e^{\frac{i PQ'}{\hbar}} \(2\cosh \frac{Q+Q'/2}{2}\)^{q/2}
\(2\cosh \frac{Q-Q'/2}{2}\)^{q/2}\\
\times \bra{Q-\frac{Q'}{2}}\biggl(2\cosh \frac{\hat{P}}{2} \biggr)^p \ket{Q+\frac{Q'}{2}}.
\ea
\ee
The last part is written as
\be
\bra{Q-\frac{Q'}{2}}\biggl(2\cosh \frac{\hat{P}}{2} \biggr)^p \ket{Q+\frac{Q'}{2}}
=\int_{-\infty}^\infty 
\frac{d P'}{2\pi } 
e^{-\frac{i P' Q'}{\hbar}}
\( 2\cosh \frac{P'}{2} \)^p.
\ee
Therefore, we find
\be
\cO_\text{W}(Q,P)=\int_{-\infty}^\infty \frac{d Q' d P'}{(2\pi)^2} 
e^{\frac{i (P-P')Q'}{2\pi}}  \(4\cosh^2 \frac{Q}{2}+4\sinh^2 \frac{k Q'}{4}\)^{q/2} \(2\cosh \frac{P'}{2}\)^p,
\ee
where we have rescaled the integration variable $Q' \to k Q'$.
As in \cite{Hatsuda}, we expand the integrand around $k=0$,
\be
\(\cosh^2 \frac{Q}{2}+\sinh^2 \frac{k Q'}{4}\)^{q/2}=\sum_{m=0}^\infty c_m(x) (k Q')^{2m}.
\ee
Then the integral over $Q'$ gives the derivative of the delta function:
\be
\int_{-\infty}^\infty \frac{d Q'}{2\pi} e^{\frac{i (P-P')Q'}{2\pi}} (Q')^{n}=(-2\pi i)^{n} \delta^{(n)}(P-P').
\ee
Thus one can easily perform the integral over $P'$
\be
\ba
\int_{-\infty}^\infty \frac{d P'}{2\pi}(2\pi i)^{2m} \delta^{(2m)}{(P-P')}\(2\cosh \frac{P'}{2}\)^p
=(2\pi i \pd_P)^{2m} \(2\cosh \frac{P}{2}\)^p .
\ea
\ee
Using these results, we finally get
\be
\ba
\cO_\text{W}(Q,P)&=\sum_{m=0}^\infty c_m(x) (2\pi i k \pd_P)^{2m} \(2\cosh \frac{P}{2}\)^p \\
&=\(4\cosh^2 \frac{Q}{2}-4\sin^2 \frac{\pi k \pd_P}{2}\)^{q/2} \(2\cosh \frac{P}{2}\)^p.
\ea
\ee

\section{Differential operators}\label{sec:diff}
Here we list the explicit forms of the differential operators $\cD^{(n)}$ up to $n=4$.
Although we have actually computed the differential operators up to $n=17$, 
it is too long to write down and we do not write the explicit forms for $n\geq 5$.
They are available upon request to the authors.
\be
\scriptsize
\ba
\cD^{(1)}&= \frac{\pi^2}{96} 
\frac{p^2 q^2 \del_\mu^2 \left(1-\del_\mu^2\right) }{(1+p\del_\mu ) (1+q\del_\mu)} ,\\
\cD^{(2)}&=\left( \frac{\pi^2}{96} \right)^2 
\frac{p^3 q^3 \del_\mu^3 \left( 1-\del_\mu^2\right)  
\left(-\frac{7}{10} p q\del_\mu^3  -\frac{12}{5}(p+q) \del_\mu^2  
+\left(-\frac{17 p  q}{10}-8\right)\del_\mu  -\frac{12 (p+q)}{5}\right)}
{( 1+p\del_\mu ) (3+p\del_\mu ) (1+q\del_\mu ) (3 +q\del_\mu )} ,\\
\cD^{(3)}&=\left( \frac{\pi^2}{96} \right)^3 
\frac{p^3 q^3 \del_\mu^3 \left(1-\del_\mu^2\right)  }
       {(1+p\del_\mu ) (3+ p\del_\mu ) (5+p\del_\mu ) (1+q\del_\mu ) (3+q\del_\mu ) (5+q\del_\mu )}
  \Biggl[\frac{31}{70} p^3 q^3 \del_\mu^7 +\frac{156}{35} p^2 q^2 (p+q)\del_\mu^6 \\
&\ \ \ \ \ +\frac{1}{35}p q \left(p^2 \left(89 q^2+464\right)+1544 p q+464 q^2\right) \del_\mu^5
+\frac{8}{35} \Bigl( p^3 \left( 93 q^2+36\right)+p^2 q \left(93 q^2+560\right) \\
&\ \ \ \ \  +560 p q^2+36 q^3\Bigr) \del_\mu^4 
+\frac{1}{70} \Bigl(p^3 q \left( 367 q^2+2272\right)+16 p^2 \left(767 q^2+336\right)+32 p q \left(71 q^2+784\right) \\
&\ \ \ \ \  +5376 q^2\Bigr) \del_\mu^3 
+\frac{4}{35} \Bigl(9 p^3 \left(23 q^2+8\right)+p^2 q \left(207 q^2+2240\right)+448 p \left(5 q^2+4\right)+8 q \left( 9q^2+224\right)\Bigr) \del_\mu^2 \\
&\ \ \ \ \  +\frac{64}{35} \left(23 p^3 q+6 p^2 \left(3 q^2+7\right)+p q \left(23 q^2+112\right)+42
   q^2+56\right) \del_\mu 
+\frac{192}{7} \left( p^3+q^3\right)\Biggr] ,\nonumber 
\ea
\ee
\be
\scriptsize
\allowdisplaybreaks[4]
\ba
\cD^{(4)}&=\left( \frac{\pi^2}{96} \right)^4
\frac{p^3 q^3 \del_\mu^3 \left(1-\del_\mu^2\right) }
{(1+p\del_\mu ) (3+p\del_\mu ) (5+p\del_\mu ) (7+p\del_\mu )  
 (1+q\del_\mu ) (3+q\del_\mu ) (5+q\del_\mu ) (7+q\del_\mu )} \\
&\ \ \ \ \  \Biggl[ -\frac{381 p^5 q^5 }{1400} \del_\mu^{11}
-\frac{942}{175} p^4 q^4 (p+q) \del_\mu^{10}
-\frac{p^3 q^3 \left(\left(3481 q^2+56128\right) p^2+147360 q p+56128 q^2\right) }{1400}\del_\mu^9
\\
&\ \ \ \ \  -\frac{6}{175} p^2 q^2 \Bigl( \left(1327 q^2+4000\right) p^3+q \left(1327 q^2+22512\right) p^2+22512 q^2 p+4000 q^3\Bigr) \del_\mu^8 \\
&\ \ \ \ \  -\frac{p q}{1400} \Bigl( \left( 14359 q^4+398528 q^2+276480\right) p^4
+960 q \left(1199 q^2+3776\right) p^3 \\
&\ \ \ \ \ +64 q^2  \left(6227 q^2+121204\right) p^2+3624960 q^3 p+276480 q^4\Bigr) \del_\mu^7
-\frac{6}{175} \Bigl( \left( 4603 q^4+18320 q^2+3200\right) p^5 \\
&\ \ \ \ \  +q \left(4603 q^4+148032 q^2+106240\right) p^4+192 q^2 \left(771 q^2+2749\right)
   p^3+16 q^3 \left(1145 q^2+32988\right) p^2 \\
&\ \ \ \ \  +106240 q^4 p+3200 q^5\Bigr) \del_\mu^6
+\frac{1}{1400} \Bigl(-q \left(27859  q^4+876992 q^2+737280\right) p^5 \\
&\ \ \ \ \  -480 \left(7123 q^4+32032 q^2+5760\right) p^4-64 q \left(13703 q^4+675244
   q^2+539136\right) p^3 \\
&\ \ \ \ \  -3072 q^2 \left(5005 q^2+26036\right) p^2-147456 q^3 \left(5 q^2+234\right) p-2764800  q^4\Bigr) \del_\mu^5 \\
&\ \ \ \ \  -\frac{6}{175} \Bigl( \left(5433 q^4+30080 q^2+3200\right) p^5+q \left(5433 q^4+276048
   q^2+267520\right) p^4 \\
&\ \ \ \ \  +48 \left(5751 q^4+39276 q^2+7840\right) p^3+64 q \left(470 q^4+29457 q^2+33984\right)  p^2 \\
&\ \ \ \ \  +256 q^2 \left(1045 q^2+8496\right) p+640 q^3 \left(5 q^2+588\right)\Bigr) \del_\mu^4
-\frac{8}{175} \Bigl( q \left(20953 q^2+14880\right) p^5\\
&\ \ \ \ \ +30 \left(423 q^4+12376 q^2+1440\right) p^4+q \left(20953 q^4+551680
   q^2+1184256\right) p^3 \\
&\ \ \ \ \  +48 \left(7735 q^4+52476 q^2+16960\right) p^2+96 q \left(155 q^4+12336 q^2+20800\right)
   p+960 q^2 \left(45 q^2+848\right)\Bigr) \del_\mu^3 \\
&\ \ \ \ \ -\frac{96}{175} \Bigl( 5 \left(977 q^2+40\right) p^5+8 q \left(201
   q^2+3560\right) p^4+24 \left(67 q^4+1980 q^2+980\right) p^3 \\
&\ \ \ \ \ +q \left( 4885 q^4+47520 q^2+139776\right) p^2+64
   \left(445 q^4+2184 q^2+1200\right) p \\
&\ \ \ \ \ +40 q \left(5 q^4+588 q^2+1920\right)\Bigr) \del_\mu^2 
 -\frac{3072}{35} \Bigl( 44 q p^5+\left(9 q^2+75\right) p^4-12 q \left(q^2-8\right) p^3 \\
&\ \ \ \ \ +3 \left(3 q^4+60 q^2+28\right) p^2+\left(44 q^5+96
   q^3+448 q\right) p+75 q^4+84 q^2+200\Bigr) \del_\mu
-2304 \left(p^5+q^5\right) \Biggr] .
\ea
\ee

\section{Derivation of the TBA functional equations for $p=2$}\label{sec:TBA-p2}
Here we derive the functional equations \eqref{eq:FR-p2}.
We start with the recursion relation \eqref{eq:recursion}.
Defining new functions by $\psi_\ell^{(j)}(x)=e^{\frac{x}{2k}} \phi_\ell^{(j)}(x)$,
then the recursion relation \eqref{eq:recursion} is rewritten as
\be
\psi_\ell^{(j)}(x)=\int_{-\infty}^\infty \frac{dy}{(2\pi k)^2} \frac{x-y}{2\sinh \frac{x-y}{2k}} \frac{\psi_{\ell-1}^{(j)}(y)}{(2\cosh \frac{y}{2})^q}.
\ee
Following the argument in \cite{KM}, these functions also satisfy the following functional relation
\be
\psi_\ell^{(j)}(x+2\pi i k)+\psi_\ell^{(j)}(x-2\pi i k)+2\psi_\ell^{(j)}(x)=\frac{1}{(2\cosh \frac{x}{2})^q} \psi_{\ell-1}^{(j)}(x).
\ee
Therefore the original functions $\phi_\ell^{(j)}(x)$ satisfy
\be
-\phi_\ell^{(j)}(x+2\pi i k)-\phi_\ell^{(j)}(x-2\pi i k)+2\phi_\ell^{(j)}(x)=\frac{1}{(2\cosh \frac{x}{2})^q} \phi_{\ell-1}^{(j)}(x).
\label{eq:phi-FR}
\ee
Let us introduce a generating functional of $\phi_\ell^{(j)}(x)$:
\be
\Phi^{(j)}(x)=\sum_{\ell=0}^\infty (-\kappa)^\ell \phi_\ell^{(j)}(x).
\ee
The functional relation \eqref{eq:phi-FR} is then written as
\be
\Phi^{(j)}(x+2\pi i k)+\Phi^{(j)}(x-2\pi i k)=2(1+t(x)) \Phi^{(j)}(x),
\label{eq:Phi-FR}
\ee
where $t(x)$ is defined by \eqref{eq:r}.
We have used the identity: $-\phi_0^{(j)}(x+2\pi i k)-\phi_0^{(j)}(x-2\pi i k)+2\phi_0^{(j)}(x)=0$ for $\phi_0^{(j)}(x)=x^j$ ($j=0,1$).
One notices that the functional relation \eqref{eq:Phi-FR} is the same form as Baxter's TQ-relation.
The functions $\Phi^{(j)}(x)$ ($j=0,1$) are two independent solutions of the TQ-relation.
A crucial fact is that these two solutions satisfy the so-called \textit{quantum Wronskian relation}:
\be
\Phi^{(0)}(x+\pi i k)\Phi^{(1)}(x-\pi i k)-\Phi^{(0)}(x-\pi i k)\Phi^{(1)}(x+\pi i k)=\text{const.}
\ee
The constant is fixed by taking the limit $\kappa \to 0$.
Since we have $\Phi^{(0)}(x)=1+\cO(\kappa)$ and $\Phi^{(1)}(x)=x+\cO(\kappa)$, one easily finds that the constant
must be $-2\pi i k$.
For later convenience, we rescale $\Phi^{(j)}(x)$ by
\be
\Phi_+(x)=\Phi^{(0)}(x),\qquad
\Phi_-(x)=-\frac{1}{\pi i k} \Phi^{(1)}(x).
\ee
Then the rescaled functions satisfy the quantum Wronskian
\be
\Phi_+(x+\pi i k)\Phi_-(x-\pi i k)-\Phi_+(x-\pi i k)\Phi_-(x+\pi i k)=2.
\label{eq:qW}
\ee
As shown below, this relation plays a crucial role in deriving \eqref{eq:FR-p2}.
Our goal is to compute the diagonal elements of the resolvent:
\be
R(x)=\frac{\rho}{1+\kappa \rho}(x,x)=\sum_{n=0}^\infty(-\kappa)^n \rho^{n+1}(x,x).
\ee
Using the formula \eqref{eq:rho-power-even}, one finds
\be
R(x)=\frac{E^2(x)}{M'(x)}\( \Phi^{(0)}(x) \pd_x \Phi^{(1)}(x)- \Phi^{(1)}(x) \pd_x \Phi^{(0)}(x) \) 
=\frac{1}{2\pi i }\frac{t(x)}{\kappa} W[ \Phi+, \Phi- ] ,
\label{eq:diag-resolvent}
\ee
where $W[f,g]=f(x)g'(x)-f'(x)g(x)$ is the standard Wronskian.

Now we derive \eqref{eq:FR-p2} from the quantum Wronskian \eqref{eq:qW}.
In the following, we use a notation, for simplicity,
\be
f^\pm=f(x\pm \pi i k).
\ee
Let us first consider the square of \eqref{eq:qW}
\be
(\Phi_+^+\Phi_-^- - \Phi_+^- \Phi_-^+)^2=4.
\ee
It is easy to see that this is equivalent to
\be
\Phi_+^+ \Phi_-^+ \Phi_+^- \Phi_-^- =\frac{1}{4}(\Phi_+^+ \Phi_-^- +\Phi_+^- \Phi_-^+)^2-1.
\ee
Introducing the functions $\xi(x)$ and $\eta(x)$ by
\be
\ba
\xi(x)&=\Phi_+(x) \Phi_-(x),\\
\eta(x)&=\frac{1}{2}\bigl(\Phi_+(x+\pi i k)\Phi_-(x-\pi i k)+\Phi_+(x-\pi i k)\Phi_-(x+\pi i k) \bigr),
\ea
\ee
then we get the first equation in \eqref{eq:FR-p2}.

Next we rewrite $\eta(x+\pi i k)+\eta(x-\pi i k)$ as
\be
\ba
\eta(x+\pi i k)+\eta(x-\pi i k)&=\frac{1}{2} \biggl[ \Phi_+(x+2\pi i k)\Phi_-(x)+\Phi_+(x) \Phi_-(x+2\pi i k) \\
&\qquad +\Phi_+(x)\Phi_-(x-2\pi i k)+\Phi_+(x-2\pi i k)\Phi_-(x) \biggr]\\
&=2(1+t(x))\xi(x),
\ea
\ee
where we have used the TQ-relation \eqref{eq:Phi-FR}.
Using $r(x)$ in \eqref{eq:r}, we get the second equation in \eqref{eq:FR-p2}.

Finally we consider the Wronskian
\be
w(x)=\frac{1}{i} W[\Phi_+,\Phi_-]=\frac{1}{i}\big(\Phi_+(x)\Phi_-'(x)-\Phi_+'(x)\Phi_-(x) \bigr).
\ee
One can see
\be
\ba
\frac{w(x+\pi i k)}{\xi(x+\pi i k)}-\frac{w(x-\pi i k)}{\xi(x-\pi i k)}
&=\frac{1}{i} \biggl[ \frac{ (\Phi_+^- \Phi_-^+)'}{\Phi_+^- \Phi_-^+}-\frac{ (\Phi_+^+ \Phi_-^-)'}{\Phi_+^+ \Phi_-^-} \biggr].
\ea
\ee
From the quantum Wronskian, we have $(\Phi_+^- \Phi_-^+)'=(\Phi_+^+ \Phi_-^-)'$.
Thus one obtains
\be
\frac{w(x+\pi i k)}{\xi(x+\pi i k)}-\frac{w(x-\pi i k)}{\xi(x-\pi i k)}
=\frac{2}{i} \frac{(\Phi_+^- \Phi_-^+)'}{\Phi_+^+ \Phi_-^+ \Phi_+^- \Phi_-^-} .
\ee
On the other hand, we have
\be
\ba
\frac{\eta'(x)}{\eta^2(x)-1}=\frac{1}{\xi(x+\pi i k)\xi(x-\pi i k)} \frac{(\Phi_+^- \Phi_-^+)'+(\Phi_+^+ \Phi_-^-)'}{2} 
=\frac{(\Phi_+^- \Phi_-^+)'}{\Phi_+^+ \Phi_-^+ \Phi_+^- \Phi_-^-} .
\ea
\ee
Therefore we find the final equation in \eqref{eq:FR-p2}.
It is easy to see the grand potential is written as \eqref{eq:dJ-p2} by using \eqref{eq:diag-resolvent}.

\section{Planar solution for $p=1$}
\label{app:planar}
In this appendix, 
we compute the free energy of the $(1,q)$-model in the 't Hooft limit
\begin{\eq}
k\rightarrow \infty ,\quad N\rightarrow \infty ,\quad 
\lambda =\frac{N}{k} ={\rm fixed},\quad q={\rm fixed} .
\end{\eq}
For $p=1$, 
the canonical partition function takes the simple form
\begin{\eq}
\left. Z(N,k) \right|_{p=1}
= \frac{1}{ 2^N N!}  \int   \frac{d^N x}{(2\pi )^N}  
       \frac{\prod_{i<j} \tanh^2{\frac{x_i -x_j}{2} } }{\prod_{i}   (2\cosh{\frac{kx_i}{2} })^{q}} ,
\end{\eq}
which is the one-parameter deformation of the $N_f$-matrix model by $k$.
If we change the variable as
\begin{\eq}
z_i = e^{x_i} ,
\end{\eq}
then we find
\begin{\eq}
\left. Z(N,k) \right|_{p=1}
= \frac{1}{  N!}  \int   \frac{d^N z}{(2\pi )^N} e^{-\sum_i V(z_i )}  
       \frac{\prod_{i<j} (z_i -z_j )^2 }{\prod_{i,j} (z_i +z_j ) } , 
\end{\eq}
where
\begin{\eq}
V(z) =q\log{(z^{k/2} +z^{-k/2})} .
\end{\eq}
In the large-$k$ limit, this potential is rewritten as
\begin{\eq}
V(z) =\pi k q \Biggl[ -\frac{\log{z}}{2\pi} 
+\frac{1}{\pi^2}\left( {\rm ImLi_2}(z) +{\rm ImLi_2}(-z) \right) +\mathcal{O}(k^{-2}) \Biggr] .
\end{\eq}
In \cite{Kashaev:2015wia}, the authors have computed
the planar free energy of the matrix model with the potential
\begin{\eq}
V_{\rm KMZ}(z)=\frac{1}{g} \Biggl[ -\frac{\log{z}}{2\pi} 
+\frac{1}{\pi^2}\left( {\rm ImLi_2}(ize^\xi ) +{\rm ImLi_2}(ize^{-\xi}) \right) 
+\mathcal{O}(k^{-2}) \Biggr] ,
\end{\eq}
by using the technique in \cite{Eynard:1995zv}.
Hence if we take $g=1/(\pi kq)$ and $\xi =\pi i /2$ in their planar solution,
then we can obtain the planar free energy of the $(1,q)$ model.
Since the ABJM case corresponds to $q=1$,
this means that 
the planar free energy of $(1,q)$ model is 
the same as the one of the ABJM model with the replacement $k\rightarrow kq$:
\begin{\eq}
\left. \log{Z(N,k)} \right|_{p=1,\rm planar}
=\left. \log{Z(N,k)} \right|_{(p,q)=1,\rm planar,k\rightarrow kq} .
\end{\eq}

Recalling that the planar free energy of the ABJM theory has 
the worldsheet instanton effect with the weight $\mathcal{O}(e^{-2\pi\sqrt{\frac{2N}{k}}})$ \cite{DMP1,Cagnazzo:2009zh},
we easily see that 
the planar free energy of the $(1,q)$ model has also the non-perturbative effect of the order
\begin{\eq}
\mathcal{O}(e^{-2\pi\sqrt{\frac{2N}{kq}}}) ,
\end{\eq} 
which is the same as the expected WS instanton effect from the gravity side.

Let us see that
this result is consistent with our result on the grand potential.
As explained in \cite{MP2,GM},
the 't Hooft limit in the grand canonical language is
\begin{\eq}
\mu\rightarrow\infty ,\quad k\rightarrow \infty ,\quad \hat{\mu}=\frac{\mu}{k}={\rm fixed.}
\end{\eq}
In this limit, we can expand the grand potential as
\begin{\eq}
\left. \mathcal{J}(\mu ,k)\right|_{p=1} =\sum_{g=0}^\infty k^{2-2g} \mathcal{J}_g (\hat{\mu},q ) ,
\end{\eq}
which should be considered as the ``genus" expansion of the grand potential.
Then the ``planar" grand potential $\mathcal{J}_0 (\hat{\mu},q)$ is
related to the planar free energy $F_0 (\lambda )$ by the Legendre transformation:
\begin{\eq}
\mathcal{J}_0 (\hat{\mu},q)
=F_0 (\lambda ) -\lambda \frac{d}{d\lambda}F_0 (\lambda ) ,\quad 
\hat{\mu}=  \frac{d}{d\lambda}F_0 (\lambda ) .
\end{\eq}
Noting the planar free energy takes the form
\begin{\eq}
F_0 (\lambda ) = q^2 f_0 (\lambda_q ) ,\quad {\rm with}\ \lambda_q =\frac{\lambda}{q} ,
\end{\eq} 
then the Legendre transformation relation becomes
\begin{\eq}
\frac{\mathcal{J}_0 (\hat{\mu},q)}{q^2}
=f_0 (\lambda_q ) -\lambda_q \frac{d}{d\lambda_q}F_0 (\lambda_q ) ,\quad 
\hat{\mu}_q = \frac{\hat{\mu}}{q}=  \frac{d}{d\lambda_q }f_0 (\lambda_q ) ,
\end{\eq}
This relation leads us to
\begin{\eq}
\mathcal{J}_0 (\hat{\mu},q) 
=q^2 \left. \mathcal{J}_0 (\hat{\mu},1) \right|_{\hat{\mu}\rightarrow \hat{\mu}_q } .
\end{\eq}
We can easily check that
the perturbative grand potential \eqref{Jpert} and 
the fist two worldsheet instanton corrections \eqref{eq:J-WS-1q}
satisfy this relation:
\begin{\eq}
\left. \mathcal{J}_{\rm pert}(\mu ,k) \right|_{p=1}
=q^2 k^2 \Biggl[ \frac{2}{\pi^2} \hat{\mu}_q^3 
+\frac{1}{24}\hat{\mu}_q +\left. A_{1,1}(k) \right|_{\mathcal{O}(k^2)}  \Biggr] +\mathcal{O}(1) ,
\end{\eq}
and
\begin{\eq}
d_1 (1,q,k) = \frac{q^2 k^2}{4\pi^2} +\mathcal{O}(1) ,\quad
d_2 (1,q,k) = -\frac{3q^2 k^2}{16\pi^2} +\mathcal{O}(1) .
\end{\eq}

\section{On the $q$-Gamma function}\label{sec:qGamma}
In this appendix we propose a useful integral representation of the $q$-gamma function.
The $q$-gamma function $\Gamma_\mathfrak{q} (z)$ is defined by
\begin{\eq}
\Gamma_\mathfrak{q} (z) =(1-\mathfrak{q})^{1-z} \frac{(\mathfrak{q};\mathfrak{q})_\infty}
{(\mathfrak{q}^z ;\mathfrak{q})_\infty} ,
\label{qgammaprod}
\end{\eq}
in terms of the $q$-Pochhammer symbol
\begin{\eq}
(a;\mathfrak{q})_\infty =\prod_{j=0}^\infty (1-a\mathfrak{q}^j ) .
\end{\eq}
The two important properties of the $q$-gamma function are
the following functional relation and the behavior in the limit $\mathfrak{q}\to1$
\begin{align}
  \Gamma_\mathfrak{q}(z+1)=
\frac{1-\mathfrak{q}^z}{1-\mathfrak{q}}\Gamma_\mathfrak{q}(z),\qquad
\lim_{\mathfrak{q}\to1} \Gamma_\mathfrak{q}(z)=\Gamma(z).
\label{qgammafunc}
\end{align}

The infinite product representation \eqref{qgammaprod} of the $q$-gamma function 
is well-defined when $|\mathfrak{q}|<1$.
However, in our case of interest $\mathfrak{q}=e^{i\hbar}$ with
$\hbar\in\mathbb{R}$, we have to deal with the $q$-gamma function
with $|\mathfrak{q}|=1$.
In this case, the naive infinite product expression
\eqref{qgammaprod} per se is ill-defined, and we have to 
define the $q$-gamma function with $|\mathfrak{q}|=1$
as a certain analytic continuation from $|\mathfrak{q}|<1$.
In the literature, such analytic continuation
was proposed by using either the double sine
function \cite{nishizawa} or the Faddeev's quantum dilogarithm integral \cite{Faddeev:1995nb}.

In this paper, we propose
an alternative integral representation of the $q$-gamma function with
$|\mathfrak{q}|=1$, which is useful for 
the numerical calculation of the instanton coefficient
in \eqref{r1qGamma}.
We regularize the infinite product 
appearing in  $q$-Pochhammer symbols 
by using the zeta-function regularization. 
For $\mathfrak{q}=e^{i\hbar}$, the $q$-Pochhammer symbols in \eqref{qgammaprod} can be rewritten as
\be
\ba
(\mathfrak{q}^z;\mathfrak{q})_\infty&=\prod_{n=0}^\infty\(-i\hbar\mathfrak{q}^{\frac{n+z}{2}} \frac{\sin\frac{\hbar(n+z)}{2}}{\frac{\hbar}{2}}\)=(-i\hbar)^{\zeta(0,z)}\mathfrak{q}^{\hf \zeta(-1,z)}
\prod_{n=0}^\infty \frac{\sin\frac{\hbar(n+z)}{2}}{\frac{\hbar}{2}},\\
(\mathfrak{q};\mathfrak{q})_\infty&=(-i\hbar)^{\zeta(0,1)}\mathfrak{q}^{\hf \zeta(-1,1)}
\prod_{n=0}^\infty \frac{\sin\frac{\hbar(n+1)}{2}}{\frac{\hbar}{2}},
\label{prodzeta}
\ea
\ee
where $\zeta(s,a)$ denotes the Hurwitz zeta function
\begin{align}
 \zeta(s,a)=\sum_{n=0}^\infty\frac{1}{(n+a)^s}.
\end{align}
Plugging  the value of $\zeta(s,a)$ at $s=0,-1$ 
\be
 \zeta(0,a)=\hf-a,\qquad \zeta(-1,a)=\hf\(\frac{1}{6}+a-a^2\) ,
\ee
into \eqref{prodzeta}, we find
\begin{align}
 \Gamma_\mathfrak{q}(z)=e^{\frac{i\hbar}{4}(z-1)(z-2)}\left(\frac{\sin\frac{\hbar}{2}}{\frac{\hbar}{2}}\right)^{1-z}
\prod_{n=0}^\infty \frac{\sin\frac{\hbar(n+1)}{2}}{\frac{\hbar}{2}}
\frac{\frac{\hbar}{2}}{\sin\frac{\hbar(n+z)}{2}} .
\label{sinprod}
\end{align}
Now, let us consider the $\hbar$ expansion of the infinite product part in \eqref{sinprod}.
Using the expansion
\begin{align}
 \log\(\frac{\sin\frac{x}{2}}{\frac{x}{2}}\)=\sum_{m=1}^\infty \frac{(-1)^mB_{2m}}{2m(2m)!}x^{2m},
\label{logsinexpand}
\end{align}
we find
\be
\ba
 &\log \prod_{n=0}^\infty \frac{\sin\frac{\hbar(n+1)}{2}}{\frac{\hbar}{2}}
\frac{\frac{\hbar}{2}}{\sin\frac{\hbar(n+z)}{2}}\\
=&\sum_{n=0}^\infty \Big[\log(n+1)-\log(n+z)\Big]+
\sum_{n=0}^\infty \sum_{m=1}^\infty 
\frac{(-1)^{m}B_{2m}\hbar^{2m}}{2m(2m)!}\Big[(n+1)^{2m}-(n+z)^{2m}\Big]\\
=&-\zeta'(0,1)+\zeta'(0,z)+\sum_{m=1}^\infty 
\frac{(-1)^{m}B_{2m}\hbar^{2m}}{2m(2m)!}
\Big[\zeta(-2m,1)-\zeta(-2m,z)\Big].
\ea
\label{logzeta}
\ee
This can be further simplified by using the relation
\begin{align}
 \zeta'(0,z)=\log\frac{\Gamma(z)}{\rt{2\pi}},\qquad
\zeta(-2m,z)=-\frac{B_{2m+1}(z)}{2m+1},
\end{align}
and \eqref{logzeta} becomes
\begin{align}
 \log \prod_{n=0}^\infty \frac{\sin\frac{\hbar(n+1)}{2}}{\frac{\hbar}{2}}
\frac{\frac{\hbar}{2}}{\sin\frac{\hbar(n+z)}{2}}=\log\Gamma(z)+
\sum_{m=1}^\infty 
\frac{(-1)^{m}B_{2m}B_{2m+1}(z)\hbar^{2m}}{2m(2m+1)!}.
\label{logsum}
\end{align}
Putting all together, we find the following
representation of the $q$-gamma function with $|\mathfrak{q}|=1$
\begin{align}
 \Gamma_\mathfrak{q}(z)=e^{\frac{i\hbar}{4}(z-1)(z-2)}\Gamma(z)
\left(\frac{\sin\frac{\hbar}{2}}{\frac{\hbar}{2}}\right)^{1-z}
\exp\left[\sum_{m=1}^\infty 
\frac{(-1)^{m}B_{2m}B_{2m+1}(z)\hbar^{2m}}{2m(2m+1)!}\right].
\label{ourqgamma}
\end{align}
Using the property of the Bernoulli polynomial \eqref{bernouli-shift},
one can show that \eqref{ourqgamma} indeed satisfies the functional relation
in \eqref{qgammafunc}, as required. Also, one can easily see that \eqref{ourqgamma} reduces to the 
usual gamma function in the limit $\hbar\to0$.

However, \eqref{ourqgamma} is still a formal expression since the summation in the exponential factor
is a divergent asymptotic series.
When $0<z<1$, we can resum this series by using the integral representation of the Bernoulli polynomial \eqref{eq:Bernoulli-int} and \eqref{logsinexpand}.
Finally, we arrive at our integral representation of the $q$-gamma function
valid for $0<z<1$ and $|\mathfrak{q}|=1$
\begin{align}
 \Gamma_\mathfrak{q}(z)=e^{\frac{i\hbar}{4}(z-1)(z-2)}\Gamma(z)
\left(\frac{\sin\frac{\hbar}{2}}{\frac{\hbar}{2}}\right)^{1-z}
\exp\left[-\int_0^\infty dt\frac{\sin2\pi z}{\cosh 2\pi t-\cos2\pi z}\log\left(\frac{\sinh\frac{\hbar t}{2}}{\frac{\hbar t}{2}}\right)\right] .
\end{align}
For the case $z\not\in(0,1)$, a similar integral representation can be obtained by using
the functional relation \eqref{qgammafunc} repeatedly.

\section{Exact values of $Z(N,k)$ and instanton corrections}\label{sec:exactZ}\label{sec:variousJ}
Using the exact values of the partition functions for various integral $(p,q,k)$,
we can determine the non-perturbative part of the modified grand potential%
\footnote{The modified grand potential $J(\mu,k)$ is related to the full grand potential $\cJ(\mu,k)$ by
\[ e^{\cJ(\mu,k)}=\sum_{n \in \mathbb{Z}} e^{J(\mu+2\pi i n,k)}.
\]
As shown in \cite{HMO2}, the modified grand potential removes the ``oscillatory part'' from the full grand potential.
}
\begin{\eq}
J_{\rm np}(\mu ,k) = J (\mu ,k) -J_{\rm pert}(\mu ,k) ,
\end{\eq}
 by the numerical fitting,
in a similar way as the ABJM case \cite{HMO2}. 

In this appendix, we list the non-perturbative part of the grand potential
$J_{\rm np}(\mu ,k)$
and the exact partition functions $Z(N,k)$ for $N=2,3,4,$
for various integral $(p,q,k)$. We drop the $N=1$ case since
we know the exact value of $Z(1,k)$ in a closed form \eqref{Z-one} for general $(p,q,k)$.
Actually we have computed the exact partition
functions for higher $N\geq5$, but they are too lengthy to write down
in this appendix.
We have also computed the exact partition functions for several other $(p,q,k)$'s 
which are not listed below. They are available upon request to the authors.

\paragraph{The case of $\bm{(p,q)=(1,2)}$.}
\begin{\eqa}
Z(2,2)&=&\frac{\pi ^2-8}{1024 \pi ^2},\quad
Z(3,2)=\frac{61 \pi ^2-600}{368640 \pi ^3},\quad
Z(4,2)=\frac{960-9424 \pi ^2+945 \pi ^4}{94371840 \pi ^4},\nn
J_{\rm np}(\mu ,2)
&=&\frac{2\mu+2}{\pi}e^{-\mu}+\left[-\frac{10\mu^2+7\mu+7/2}{\pi^2}+1\right]e^{-2\mu}
+\frac{88\mu+52/3}{3\pi}e^{-3\mu}\nn
&&+\left[-\frac{269\mu^2+193\mu/4+265/16}{\pi^2}+58\right]e^{-4\mu}+
\frac{4792\mu+1102/5}{5\pi}e^{-5\mu} ,\nn
Z(2,3)&=&\frac{89 \pi ^2-864}{31104 \pi ^2},\quad
Z(3,3)=\frac{-21384+13311 \pi ^2-2048 \sqrt{3} \pi ^3}{10077696 \pi ^3},
\nn
Z(4,3)&=&\frac{614304-1821312 \pi ^2-32768 \sqrt{3} \pi ^3+196297 \pi ^4}{1934917632 \pi ^4},\nn
J_{\rm np}(\mu ,3)&=& 
\frac{8}{3}e^{-\frac{2\mu}{3}}-6e^{-\frac{4\mu}{3}}+\left[-\frac{4\mu^2+2\mu+1}{3\pi^2}+\frac{88}{9}\right]e^{-2\mu}-\frac{238}{9}e^{-\frac{8\mu}{3}}\nn
&&\quad+\frac{848}{15}e^{-\frac{10\mu}{3}}+\left[-\frac{52\mu^2+\mu+9/4}{6\pi^2}-\frac{1540}{9}\right]e^{-4\mu}
+\frac{82672}{189}e^{-\frac{14\mu}{3}},\nn
Z(2,4)&=&\frac{5 \pi ^2-48}{8192 \pi ^2},\quad
Z(3,4)=\frac{-2640+833 \pi ^2-180 \pi ^3}{5898240 \pi ^3},\nn
Z(4,4)&=&\frac{6400-15776 \pi ^2-4864 \pi ^3+3081 \pi ^4}{402653184 \pi ^4},\nn
J_{\rm np}(\mu ,4)
&=& 2\rt{2}e^{-\frac{\mu}{2}}+\left[\frac{\mu+1}{\pi}-4\right]e^{-\mu}
+\frac{16\rt{2}}{3}e^{-\frac{3\mu}{2}}
+\left[-\frac{10\mu^2+7\mu+7/2}{2\pi^2}-\frac{45}{2}\right]e^{-2\mu}\nn
&&\quad+\frac{288\rt{2}}{5}e^{-\frac{5\mu}{2}}
+\left[\frac{88\mu+52/3}{6\pi}-\frac{640}{3}\right]e^{-3\mu},\nn
Z(2,6)&=&\frac{331 \pi ^2-3240}{746496 \pi ^2},\quad
Z(3,6)=\frac{-495720+287037 \pi ^2-43520 \sqrt{3} \pi ^3}{2418647040 \pi ^3},\nn
Z(4,6)&=&\frac{459794880-1161396144 \pi ^2-320716800 \sqrt{3} \pi ^3+289774225 \pi ^4}{50153065021440 \pi ^4},\nn
J_{\rm np}(\mu ,6)
&=& \frac{8}{\rt{3}}e^{-\frac{\mu}{3}}-\frac{14}{3}e^{-\frac{2\mu}{3}}
+\left[\frac{2(\mu+1)}{3\pi}+8\rt{3}\right]e^{-\mu}-\frac{154}{3}e^{-\frac{4\mu}{3}} .
\end{\eqa}

\paragraph{The case of $\bm{(p,q)=(1,3)}$.}
\begin{\eqa}
Z(2,2)&=&\frac{32-3 \pi ^2}{12288 \pi ^2},\quad
Z(3,2)=\frac{7552-765 \pi ^2}{47185920 \pi ^2},\quad
Z(4,2)=\frac{143360-278784 \pi ^2+26775 \pi ^4}{42278584320 \pi ^4},\nn
J_{\rm np}(\mu ,2)
&=& \frac{2(2\mu+3)}{\rt{3}\pi}e^{-\frac{2\mu}{3}}+\left[-\frac{(2\mu+3)^2}{\pi^2}+\frac{2}{3}\right]e^{-\frac{4\mu}{3}}\nn
&&+
\left[\frac{(2\mu+3)^3}{\rt{3}\pi^3}+\frac{76\mu^2}{3\pi^2}
+\frac{47(2\mu+1)}{6\pi^2}+\frac{2(2\mu+3)}{\rt{3}\pi}-8\right]e^{-2\mu}\nn
&&+\left[-\frac{(2\mu+3)^4}{2\pi^4}-\frac{2(2\mu+3)^2}{\pi^2}
-\frac{166\mu+133/4}{\rt{3}\pi}+\frac{2}{3}\right]e^{-\frac{8\mu}{3}}\nn
&&+\left[\frac{\rt{3}(2\mu+3)^5}{5\pi^5}+\frac{8(2\mu+3)^3}{\rt{3}\pi^3}+\frac{332\mu^2+1129\mu/2+399/4}{\pi^2}-\frac{4\rt{3}(2\mu+3)}{\pi}-48\right]e^{-\frac{10\mu}{3}} ,\nn
Z(2,4)&=&\frac{105 \pi ^2-1024}{786432 \pi ^2},\quad
Z(3,4)=\frac{-1024-2624 \pi +939 \pi ^2}{100663296 \pi ^2},\nn
Z(4,4)&=&\frac{367001600-871395328 \pi ^2-351375360 \pi ^3+196369425 \pi ^4}{432932703436800 \pi ^4},\nn
J_{\rm np}(\mu ,4)
&=&6e^{-\frac{\mu}{3}}+\left[\frac{2(2\mu+3)}{\rt{3}\pi}-11\right]e^{-\frac{2\mu}{3}}+
\left[-\frac{3\rt{3}(2\mu+3)}{\pi}+35\right]e^{-\mu} .
\end{\eqa}

\paragraph{The case of $\bm{(p,q)=(1,4)}$.}
\begin{\eqa}
Z(2,2)&=&\frac{27 \pi ^2-256}{393216 \pi ^2},\quad
Z(3,2)=\frac{21975 \pi ^2-216832}{13079937024 \pi ^3},\nn
Z(4,2)&=&\frac{3153920000-7092337152 \pi ^2+686225925 \pi ^4}{16072626615091200 \pi ^4},\nn
J_{\rm np}(\mu ,2)
&=& \frac{\rt{2}(2\mu+4)}{\pi}e^{-\frac{\mu}{2}}+\left[-\frac{(2\mu+4)^2}{\pi^2}+\frac{3(\mu+1)}{\pi}\right]e^{-\mu}+\left[\frac{2\rt{2}(2\mu+4)^3}{3\pi^3}-4\rt{2}\right]e^{-\frac{3\mu}{2}} ,\nn
Z(2,3)&=&\frac{179155 \pi ^2-1767096}{181398528 \pi ^2},\quad
Z(3,3)=\frac{-585293688+908028549 \pi ^2-155975680 \sqrt{3} \pi ^3}{7405413507072 \pi ^3},\nn
Z(4,3)&=&\frac{1168603329600-2706064255728 \pi ^2+120082923520 \sqrt{3} \pi ^3+195979586775 \pi ^4}{28436787867156480 \pi ^4},\nn
J_{\rm np}(\mu ,3)
&=&\frac{16}{\rt{3}}e^{-\frac{\mu}{3}}-\frac{86}{3}e^{-\frac{2\mu}{3}}
+\left[-\frac{2\mu+2}{3\pi}+88\rt{3}\right]e^{-\mu}
-\frac{2458}{3}e^{-\frac{4\mu}{3}},\nn
Z(2,4)&=&\frac{5841 \pi ^2-57344}{150994944 \pi ^2},\quad
Z(3,4)=\frac{-14508032+5682711 \pi ^2-1340955 \pi ^3}{3348463878144 \pi ^3},\nn
Z(4,4)&=&\frac{1192021196800-2569664274432 \pi ^2-819920683008 \pi ^3+509113016685 \pi ^4}{16757612738468904960 \pi ^4},\nn
J_{\rm np}(\mu ,4)
&=&\frac{4}{\sin\frac{\pi}{8}}e^{-\frac{\mu}{4}}+\left[\frac{\rt{2}(\mu+2)}{\pi\tan\frac{\pi}{8}}+\text{const}\right]e^{-\frac{\mu}{2}} .
\end{\eqa}

\paragraph{The case of $\bm{(p,q)=(1,6)}$.}
\begin{\eqa}
Z(2,2)&=&\frac{3375 \pi ^2-32768}{754974720 \pi ^2},\quad
Z(3,2)=\frac{137216175 \pi ^2-1354203136}{6607236759552000 \pi ^3},\nn
Z(4,2)&=&\frac{18673845640626176-36455447012966400 \pi ^2+3502003677046875 \pi ^4}{20368878283608953978880000 \pi ^4},\nn
J_{\rm np}(\mu ,2)
&=&\frac{2(2\mu+6)}{\pi}e^{-\frac{\mu}{3}}+\left[-\frac{(2\mu+6)^2}{\pi^2}
+\frac{\rt{3}(4\mu+6)}{\pi}-2\right]e^{-\frac{2\mu}{3}}\nn
&&+\left[\frac{(2\mu+6)^3}{\pi^3}-
\frac{3\rt{3}(2\mu+6)(4\mu+6)}{2\pi^2}+\frac{32\mu+56}{3\pi}-\frac{8}{\rt{3}}\right]e^{-2\mu}  .
\end{\eqa}

\paragraph{The case of $\bm{(p,q)=(2,3)}$.}
\begin{\eqa}
Z(2,1)&=&\frac{3 \left(5 \pi ^2-48\right)}{40960 \pi ^2},\quad
Z(3,1)=
\frac{6784-687 \pi ^2}{70778880 \pi ^3},\nn
Z(4,1)&=&\frac{45731840-92351928 \pi ^2+8887725 \pi ^4}{10463949619200 \pi ^4},
\nn
J_{\rm np}(\mu ,1)
&=&\frac{4(2\mu+3)}{\pi}e^{-\frac{2\mu}{3}}
+\left[-\frac{(2\mu+3)^2}{\pi^2}-\frac{3\rt{3}(4\mu+3)}{2\pi}
-\frac{10}{3}\right]e^{-\frac{4\mu}{3}} \nn
&&+\left[\frac{2(2\mu+3)^3}{\rt{3}\pi^3}+\frac{8(4\mu^2+49\mu/2+49/4)}{3\pi^2}
+\frac{4(2\mu+3)}{\rt{3}\pi}+16\right]e^{-2\mu},\nn
Z(2,2)&=&\frac{45 \pi ^2-436}{983040 \pi ^2},\quad
Z(3,2)=\frac{-240+1870 \pi ^2-187 \pi ^4}{235929600 \pi ^5},\nn
Z(4,2)&=&\frac{887040000-2063550720 \pi ^2+6724822720 \pi ^4-8578838736 \pi ^6+802234125 \pi ^8}{23439247147008000 \pi ^8},\nn
J_\text{np}(\mu ,2)
&=&4\rt{3}e^{-\frac{\mu}{3}}+\left[\frac{\rt{3}(2\mu+3)}{\pi}-\frac{37}{3}\right]e^{-\frac{2\mu}{3}}+\left[-\frac{8\mu+20}{\pi}+22\rt{3}\right]e^{-\mu}\nn
&&+\left[-\frac{3\mu^2+9\mu-109/3}{\pi^2}+\frac{67\mu+17}{\rt{3}\pi}-\frac{391}{3}\right]e^{-\frac{4\mu}{3}} ,\nn
Z(2,4)&=&\frac{-6076-2160 \pi +1305 \pi ^2}{31457280 \pi ^2},\nn
Z(3,4)&=&\frac{-15360-57600 \pi +56560 \pi ^2+956280 \pi ^3+431972 \pi ^4-235575 \pi ^5}{483183820800 \pi ^5},\nn
Z(4,4)&=&(7096320000-33621181440 \pi ^2-49550054400 \pi ^3+238227230780 \pi ^4\nn
&&+1442439028800 \pi ^5-1892035274352 \pi
   ^6-2049910590420 \pi ^7\nn
&&+795439693125 \pi ^8)
/(24001789078536192000 \pi ^8),\nn
J_\text{np}(\mu ,4)&=&
12\rt{2}e^{-\frac{\mu}{6}}+(-26+5\rt{3})e^{-\frac{\mu}{3}}.
\end{\eqa}

\paragraph{The case of $\bm{(p,q)=(2,4)}$.}
\begin{\eqa}
Z(2,1)&=&\frac{105-4 \pi ^2}{483840 \pi ^4},\quad
Z(3,1)=\frac{-10395+9900 \pi ^2-896 \pi ^4}{638668800 \pi ^6},\nn
Z(4,1)&=&\frac{-154729575+295214920 \pi ^2-120054480 \pi ^4+9294336 \pi ^6}{334764638208000 \pi ^8},
\nn
J_{\rm np}(\mu ,1)
&=&\frac{4(\mu+2)}{\pi}e^{-\frac{\mu}{2}}
+\left[\frac{2\mu^2-18\mu-18}{\pi^2}-2\right]e^{-\mu}+ 
\left[\frac{32(\mu+2)^3}{3\pi^3}+\frac{-128\mu+80/3}{3\pi}\right]e^{-\frac{3\mu}{2}}\nn
&&+\left[-\frac{64(\mu+2)^4}{3\pi^4}+\frac{-295\mu^2/3+3767\mu/18-4945/36}{\pi^2}+82\right]e^{-2\mu},\nn
Z(2,2)&=&\frac{134400-96496 \pi ^2+8505 \pi ^4}{990904320 \pi ^4},\nn
Z(3,2)&=&\frac{-9461760-11623040 \pi ^2+55137032 \pi ^4-5457375 \pi ^6}{41855798476800 \pi ^6},\nn
Z(4,2)&=&(-6125543424000+15615912632320 \pi ^2+7125815457280 \pi ^4\nn
&&-13725449056224 \pi ^6+1301927752125 \pi
   ^8)
/(702052330547183616000 \pi ^8),\nn
J_{\rm np}(\mu ,2)
&=&8\rt{2}e^{-\frac{\mu}{4}}+\left[\frac{4(\mu+2)}{\pi}-22\right]e^{-\frac{\mu}{2}}+\left[-\frac{16\rt{2}(3\mu+4)}{3\pi}+\frac{373}{4}\right]e^{-\frac{3\mu}{4}} ,\nn
Z(2,3)&=&\frac{6200145-5356836 \pi ^2-752640 \sqrt{3} \pi ^3+896000 \pi ^4}{85710804480 \pi ^4},\nn
Z(3,3)&=&(237546155385-795888797220 \pi ^2-323485747200 \sqrt{3} \pi ^3\nn
&&+1758863922048 \pi ^4+353173708800 \sqrt{3} \pi
   ^5-346931200000 \pi ^6)
/(2226900409245388800 \pi ^6),\nn
Z(4,3)&=&(-60294434727802275+232273233091940040 \pi ^2-133698981198182400 \sqrt{3} \pi ^3\nn
&&+498643327433450160 \pi
   ^4+393619532866560000 \sqrt{3} \pi ^5-1023964661186984448 \pi ^6\nn
&&-238080482537963520 \sqrt{3} \pi ^7+208424280064000000
   \pi ^8)
/(94547421599315102466048000 \pi ^8),\nn
J_{\rm np}(\mu ,3)
&=&\frac{32}{\rt{3}}e^{-\frac{\mu}{6}}-16e^{-\frac{\mu}{3}}.
\end{\eqa}

\paragraph{The case of $\bm{(p,q)=(2,6)}$.}
\begin{\eqa}
Z(2,1)&=&\frac{93 \pi ^2-770}{39916800 \pi ^4},\quad
Z(3,1)=\frac{-18293275+12655344 \pi ^2-1094400 \pi ^4}{38109367296000 \pi ^6},\nn
Z(4,1)&=&\frac{-39855355314775+64469243716878 \pi ^2-24750110932272 \pi ^4+1887326438400 \pi ^6}{14772580993648558080000 \pi ^8},\nn
J_{\rm np}(\mu ,1)
&=&\frac{8(\mu+3)}{\rt{3}\pi}e^{-\frac{\mu}{3}}
+\left[-\frac{4(\mu+3)^2}{\pi^2}+\frac{3\rt{3}(2\mu+3)}{\pi}-\frac{10}{3}\right]e^{-\frac{2\mu}{3}}\nn
&&+\left[\frac{16(\mu+3)^3}{\rt{3}\pi^3}-\frac{8(4\mu^2+49\mu+49)}{3\pi^2}
+\frac{8(\mu+3)}{\rt{3}\pi}-16\right]e^{-\mu} ,\nn
Z(2,2)&=&\frac{40370176-118938552 \pi ^2+11694375 \pi ^4}{20927899238400 \pi ^4},\nn
Z(3,2)&=&\frac{-51211183063040+7706727614976 \pi ^2+52775122747500 \pi ^4-5373085843125 \pi ^6}{2557476346993311744000 \pi ^6},\nn
Z(4,2)&=&(-25823407803062655385600+50663329067554514927616 \pi ^2\nn
&&+18648795169075900287744 \pi ^4
-39645716122443476152080 \pi^6\nn
&&+3775526179656379273125 \pi ^8)
/(507582017961853925369252413440000 \pi ^8),\nn
J_{\rm np}(\mu ,2)
&=& 24e^{-\frac{\mu}{6}}+\left[\frac{10(\mu+3)}{\rt{3}\pi}-50\right]e^{-\frac{\mu}{3}} .
\end{\eqa}

\paragraph{The case of $\bm{(p,q)=(3,4)}$.}
\begin{\eqa}
Z(2,1)&=&\frac{-2880+13680 \pi ^2-13336 \pi ^4+1215 \pi ^6}{8847360 \pi ^6},\nn
Z(3,1)&=&\frac{241920+1028160 \pi ^2-3631824 \pi ^4+2554832 \pi ^6-222669 \pi ^8}{107017666560 \pi ^9},\nn
Z(4,1)&=&(-7770470400-27076896000 \pi ^2-567425295360 \pi ^4+1304054950880 \pi ^6\nn
&&-764702084232 \pi ^8+64686265875 \pi
   ^{10})
/(843812897292288000 \pi ^{10}),\nn
J_{\rm np}(\mu ,1)
&=&8\rt{3}e^{-\frac{\mu}{3}}+\left[\frac{2\rt{3}(2\mu+3)}{\pi}-\frac{915}{18}\right]e^{-\frac{2\mu}{3}}
+\left[-\frac{28\mu+34}{\pi}+161\rt{3}\right]e^{-\mu} ,\nn
Z(2,2)&=&\frac{-241920+781200 \pi ^2-1331896 \pi ^4+127575 \pi ^6}{39636172800 \pi ^6},\nn
Z(3,2)&=&\frac{-10866240-66224400 \pi ^2+102942268 \pi ^4-9739065 \pi ^6}{669692775628800 \pi ^7},\nn
Z(4,2)&=&(16428264652800-265480498483200 \pi ^2+970866947930880 \pi ^4\nn
&&-3215125817496320 \pi ^6+5696789218414112 \pi
   ^8-3547297149534048 \pi ^{10}\nn
&&+304177955956875 \pi ^{12})
/(2096796293900921733120000 \pi ^{12}),\nn
J_{\rm np}(\mu ,2)
&=&8\rt{6}e^{-\frac{\mu}{6}}+\left(-\frac{77}{3}+6\rt{3}\right)e^{-\frac{\mu}{3}}.
\end{\eqa}

\paragraph{The case of $\bm{(p,q)=(4,4)}$.}
\begin{\eqa}
Z(2,1)&=&\frac{-7875+39690 \pi ^2-23275 \pi ^4+2032 \pi ^6}{406425600 \pi ^8},\nn
Z(3,1)&=&\frac{675675-1143450 \pi ^2+17037405 \pi ^4-25454770 \pi ^6+9954912 \pi ^8-764928 \pi ^{10}}{21244678963200 \pi ^{12}},\nn
Z(4,1)&=&(82084377375+10578900231900 \pi ^2-9075519302850 \pi ^4+84295306635540 \pi ^6\nn
&&-264554733287845 \pi ^8+226822274210896
   \pi ^{10}-65334385524096 \pi ^{12}\nn
&&+4557585014784 \pi ^{14})
/(30400217826207989760000 \pi ^{16}),\nn
J_{\rm np}(\mu ,1)
&=&16e^{-\frac{\mu}{4}}+\left[\frac{12\mu+28}{\pi}-48\right]e^{-\frac{\mu}{2}}
+\left[-\frac{64(\mu+3)}{\pi}+\frac{736}{3}\right]e^{-\frac{3\mu}{4}}\nn
&&+\left[-\frac{19\mu^2+127\mu}{\pi^2}+\frac{512\mu}{\pi}+\text{const}\right]e^{-\mu} .
\end{\eqa}

\paragraph{The case of $\bm{(p,q)=(4,6)}$.}
\begin{\eqa}
Z(2,1)&=&\frac{-872025+2110185 \pi ^2-1034880 \pi ^4+84928 \pi ^6}{178827264000 \pi ^8},\nn
Z(3,1)&=&(1355404050-18596603025 \pi ^2+84994485030 \pi ^4-87718992920 \pi ^6\nn
&&+28336344192 \pi ^8-2057011200 \pi
   ^{10})
/(3073139378749440000 \pi ^{12}),\nn
Z(4,1)&=&(-25947408325460997375+98059797050517051750 \pi ^2-159596431741765643625 \pi ^4\nn
&&+1367339128216834172400 \pi
   ^6-2744252481476517451440 \pi ^8\nn
&&+1878037646578895401600 \pi ^{10}-476496174210114060288 \pi ^{12}\nn
&&+31711217482019635200
   \pi ^{14})
/(47698087690365901731790848000000 \pi ^{16}),\nn
J_{\rm np}(\mu ,1)
&=&16\rt{3}e^{-\frac{\mu}{6}}
+\left[\frac{4(\mu+3)}{\rt{3}\pi}-50\right]e^{-\frac{\mu}{3}} .
\end{\eqa}

\paragraph{The case of $\bm{(p,q,k)=(1,q,2)}$.}
In the case of $(p,q,k)=(1,q,2)$ with general $q$, from the numerical fitting we find that
the non-perturbative part of grand potential takes the 
similar form as the $N_f$-matrix model \cite{HO}
\begin{align}
 J_\text{np}(\mu,k=2)=\sum_{m=1}^\infty\mathcal{P}_m(\mu,q)e^{-\frac{2m\mu}{q}}+
\sum_{\ell=1}^\infty \Big[b_\ell(q)\mu+c_\ell(q)\Big]e^{-2\ell\mu}+\cdots,
\label{1q2}
\end{align} 
where $\mathcal{P}_m(\mu,q)$ is a $m^\text{th}$ order polynomial of $\mu$ and the ellipses
denote the contributions of bound states.
The first three terms of $\mathcal{P}_m(\mu,q)$ are given by
\begin{align}
 \mathcal{P}_1(\mu,q)&=\frac{1}{s_1}P_1,\quad
 \mathcal{P}_2(\mu,q)=-P_1^2+\frac{3s_3}{4s_1s_2}P_2-\frac{s_4}{2s_1^2s_2},\nn
\mathcal{P}_3(\mu,q)&=\frac{2s_2^2}{3s_1}P_1^3-\frac{3s_4}{2s_1}P_1P_2+\frac{10s_4s_5}{9s_1s_2s_3}P_3
+\frac{s_4^2}{s_1s_2^2}P_1+\frac{2s_6}{s_1^2s_3}-\frac{2s_4s_5}{s_1^2s_2^2},
\label{k2WS}
\end{align}
where we defined
\begin{align}
 s_n=\sin\frac{\pi n}{q},\quad
P_n=\frac{2n\mu+q}{\pi}.
\end{align}
We also conjecture that the 1-instanton coefficients $b_1(q),c_1(q)$ in \eqref{1q2} 
are given by
\begin{align}
b_1(q)=2c_1(q)=-\frac{1}{2\pi^2}\frac{\Gamma^2(-q)}{\cos(\pi q)\Gamma(-2q)}.
\label{k2conjecture}
\end{align}
By taking the limit $q\to n~(n\in\mathbb{Z})$, one can check that the conjectured form of instanton coefficients 
\eqref{k2WS} and \eqref{k2conjecture} correctly reproduces the result of 
$J_\text{np}(\mu,2)$ listed above
for the $(p,q,k)=(1,q,2)$ case with various integer $q$.

One can derive the expression of $b_1(q)$ in \eqref{k2conjecture}
by taking the limit $k\to2$ of $\gamma_1(q,k)$ given by \eqref{D1qk}.
However, if we take the limit naively, we get a wrong answer.
To reproduce
\eqref{k2conjecture}, we have to first rewrite $\gamma_1(q,k)$
by using the transformation of hypergeometric function as
\begin{align}
 \gamma_1(q,k)&=-\frac{1}{\pi^2 k}\frac{\Gamma^2(-q)}{\Gamma(-2q)}
 {}_2F_1\(-\frac{q}{2},-\frac{q}{2},\hf-q;\sin^2\frac{\pi k}{2}\)\nn
 &=
 -\frac{1}{\pi^2 k}\frac{\Gamma^2(-q)}{\Gamma(-2q)}
 {}_2F_1\(-q,-q,\hf-q;\sin^2\frac{\pi k}{4}\).
\end{align}
Taking the limit $k\to2$ in the last expression, we correctly obtain
$b_1(q)$ in \eqref{k2conjecture}.

\end{document}